\documentclass[sigconf]{acmart}
\setcopyright{none}
\renewcommand\footnotetextcopyrightpermission[1]{}

\usepackage{multirow}
\usepackage{tabularx}
\usepackage{booktabs}
\usepackage{hyperref}
\usepackage{array}
\usepackage{xcolor}
\definecolor{ricewhite}{rgb}{0.97, 0.96, 0.93}
\definecolor{verylightgray}{gray}{0.97}
\usepackage[normalem]{ulem}

\AtBeginDocument{%
  }

\setcopyright{acmlicensed}
\copyrightyear{2026}
\acmYear{2026}
\acmDOI{XXXXXXX.XXXXXXX}
\acmConference[Conference acronym 'XX]{Make sure to enter the correct
  conference title from your rights confirmation email}{June 03--05,
  2018}{Woodstock, NY}
\acmISBN{978-1-4503-XXXX-X/2018/06}

\begin{document}

\title{SQuTR: A Robustness Benchmark for Spoken Query to Text Retrieval under Acoustic Noise}

\author{Yuejie Li}
\authornotemark[1]
\affiliation{%
  \institution{Ant Group}
  \city{Hangzhou}
  \country{China}
}
\email{li_yuejie@alumni.hust.edu.cn}

\author{Ke Yang}
\authornotemark[1]
\affiliation{
  \institution{Ant Group}
  \city{Hangzhou}
  \country{China}
}
\email{yangkeai@connect.hku.hk}

\author{Yueying Hua}
\authornotemark[1]
\affiliation{%
  \institution{Soochow University}
  \country{Suzhou}
  \country{China}
}
\email{yyhua1224@outlook.com}

\author{Berlin Chen}
\authornote{Both authors contributed equally to this research.}
\affiliation{%
  \institution{University of Science and Technology of China}
  \city{Hefei}
  \country{China}
}
\email{berlin@mail.ustc.edu.cn}

\author{Jianhao Nie}
\affiliation{%
  \institution{Wuhan University}
  \city{Wuhan}
  \country{China}
}
\email{2017282110385@whu.edu.cn}

\author{Yueping He}
\affiliation{%
  \institution{Tsinghua University}
  \city{Beijing}
  \country{China}
}
\email{heyp21@mails.tsinghua.edu.cn}

\author{Caixin Kang}
\authornote{Corresponding Author}
\affiliation{%
  \institution{The University of Tokyo}
  \city{Tokyo}
  \country{Japan}
}
\email{cxkang@iis.u-tokyo.ac.jp}

\renewcommand{\shortauthors}{Li et al.}




\begin{abstract}
Spoken query retrieval is an important interaction mode in modern information retrieval.
However, existing evaluation datasets  are often limited to simple queries under constrained noise conditions, making them inadequate for assessing the robustness of spoken query retrieval systems under complex acoustic perturbations.
To address this limitation, we present SQuTR, a robustness benchmark for spoken query retrieval that includes a large-scale dataset and a unified evaluation protocol. 
SQuTR aggregates 37,317 unique queries from six commonly used English and Chinese text retrieval datasets, spanning multiple domains and diverse query types.
We synthesize speech using voice profiles from 200 real speakers and mix 17 categories of real-world environmental noise under controlled SNR levels, enabling reproducible robustness evaluation from quiet to highly noisy conditions.
Under the unified protocol, we conduct large-scale evaluations on representative cascaded and end-to-end retrieval systems.
Experimental results show that retrieval performance decreases as noise increases, with substantially different drops across systems. Even large-scale retrieval models struggle under extreme noise, indicating that robustness remains a critical bottleneck.
Overall, SQuTR provides a reproducible testbed for benchmarking and diagnostic analysis, and facilitates future research on robustness in spoken query to text retrieval.

\end{abstract}




\maketitle

\section{Introduction}

Speech has become an increasingly important interaction modality for information retrieval systems~\cite{van2023modeling,lin2024speechdpr}. From voice assistants in smart home devices to in-vehicle infotainment systems~\cite{wang2025voiceassistant}, users are turning to spoken queries as a primary means of accessing information. Unlike clear text input, spoken queries in real-world scenarios are often affected by background noise, environmental interference, and speaker variability. These factors degrade Automatic Speech Recognition (ASR) performance, and the resulting transcription errors propagate to the retriever, often causing substantial drops in retrieval effectiveness.


Although prior studies~\cite{Min2025,sidiropoulos2022impact,sidiropoulos2024multimodaldenseretrievalapproach} have noted that ASR errors can negatively affect downstream tasks, existing evaluation practices remain highly fragmented.
Speech-related work~\cite{shah2024speechrobustbenchrobustness,Park_2019,watanabe2018espnetendtoendspeechprocessing} primarily measures ASR robustness under noise using transcription-level metrics such as word error rate (WER), treating ASR as an isolated component. 
In contrast, information retrieval (IR) research~\cite{thakur2021beir,nguyen2016ms} evaluates models under the assumption that queries are clean and unambiguous text.
As a result, most IR benchmarks~\cite{muennighoff2023mteb,thakur2021beir} are constructed exclusively with text queries, while speech benchmarks~\cite{panayotov2015librispeech,radford2023robust,barker2018fifth} focus primarily on ASR performance rather than downstream retrieval performance. 
This mismatch makes it hard to compare spoken query retrieval systems under a unified, controlled protocol, and it limits our understanding of system robustness under diverse acoustic conditions.

Recent efforts have started to incorporate speech into benchmark-driven evaluation. MSEB~\cite{heigold2025massive} has introduced a retrieval-related subset with simple voice questions (SVQ) recorded under multiple environments. 
However, SVQ exhibits several limitations for robustness evaluation. First, its queries are primarily single-hop and fact-oriented, with relatively standardized question forms. Second, the associated corpora largely consist of short passages in general domains (e.g., Wikipedia), limiting contextual and task complexity. Third, although recordings are collected under several environmental conditions, noise intensity is not explicitly controlled across graded signal-to-noise ratio (SNR) levels. 

To address these limitations, we introduce SQuTR, a benchmark for spoken query to text retrieval under controlled acoustic noise. SQuTR provides (i) a large-scale spoken-query dataset derived from text-based IR benchmarks and (ii) a unified evaluation protocol that measures retrieval performance under multiple acoustic conditions.
SQuTR constructs spoken queries from real queries in six widely used English and Chinese retrieval benchmarks: FiQA-2018 (finance)~\cite{de2018inf}, HotpotQA (multi-hop QA)~\cite{yang2018hotpotqa}, Natural Questions (open-domain QA)~\cite{kwiatkowski2019natural}, MedicalRetrieval (medical)~\cite{long2022multicprmultidomainchinese}, DuRetrieval (general-domain retrieval)~\cite{qiu2022dureader-retrieval}, and T2Retrieval (passage retrieval)~\cite{xie2023t2ranking}. This design preserves realistic query complexity and broad task diversity beyond simple question templates.

We synthesize speech using CosyVoice-3~\cite{du2025cosyvoice3inthewildspeech} with 200 diverse speakers. To simulate diverse acoustic environments, we inject real-world environmental noise sampled from over a dozen real-world categories, and construct four graded acoustic conditions (Clean, Low Noise, Medium Noise, and High Noise) by controlling the signal-to-noise ratio (SNR). This design enables controlled and reproducible analysis of retrieval robustness under graded acoustic perturbations.


Our experiments systematically evaluate retrieval performance across different acoustic conditions, multiple ASR front-ends, and both lexical and dense retrievers. 
We observe a consistent decline in retrieval effectiveness as acoustic noise increases, and this degradation pattern remains stable across ASR models of varying sizes and architectures. 
These findings underscore robustness as a key challenge for practical spoken query to text retrieval.
The code and dataset are publicly available at: \textcolor{cyan}{\textit{\url{https://github.com/ttoyekk1a/SQuTR-Spoken-Query-to-Text-Retrieval}}}

\section{Related Work}

\subsection{Spoken Query Retrieval}
Spoken Query Retrieval investigates retrieving relevant information in response to spoken user queries. Prior work largely follows two paradigms: \emph{cascaded} systems~\cite{sidiropoulos2022impact} and \emph{end-to-end} speech retrieval~\cite{elizalde2022claplearningaudioconcepts,xu2025omni}.
Cascaded systems first transcribe speech with ASR and then apply mature text retrievers, such as lexical methods (e.g., BM25~\cite{robertson2009probabilistic}) or neural retrievers, benefiting from modularity and reuse of IR infrastructure. However, retrieval quality is inherently affected by transcription errors.
End-to-end approaches aim to bypass explicit transcription by directly mapping speech to retrieval representations~\cite{elizalde2022claplearningaudioconcepts,lee2015spoken}.
Despite progress, existing evaluations rarely provide controlled and systematic analysis of retrieval robustness under varying acoustic noise, making cross-system comparison under realistic conditions difficult.

\subsection{ASR Robustness and Evaluation}
Noise robustness has long been a central topic in ASR, driven by benchmarks and challenge datasets such as CHiME-style evaluations~\cite{barker2015third, barker2018fifth}. ASR robustness is typically measured by transcription-level metrics, most notably word error rate (WER) or character error rate (CER), and recent robustness benchmarks~\cite{shah2024speechrobustbenchrobustness, kim2025does} further standardize such evaluation under diverse corruptions. However, these benchmarks largely stop at transcription and do not quantify how recognition errors propagate to downstream tasks such as retrieval, limiting end-to-end robustness assessment for spoken-query retrieval systems.

\subsection{Information Retrieval Benchmarks}
Progress in IR has been closely tied to standardized benchmarks. Large-scale datasets such as DuRetrieval~\cite{qiu2022dureader-retrieval}, T2Ranking~\cite{xie2023t2ranking}, and Natural Questions~\cite{kwiatkowski2019natural}, as well as evaluation suites such as BEIR~\cite{thakur2021beir} and MTEB~\cite{muennighoff2023mteb}, have accelerated the development and comparison of retrieval models. However, these benchmarks are fundamentally \emph{text-only}: they assume clean textual queries and therefore do not capture uncertainty introduced by spoken input and acoustic variability.


Recent efforts have started to incorporate speech into benchmark-driven evaluation. Massive Sound Embedding Benchmark (MSEB) introduces Simple Voice Questions (SVQ), a large collection of short spoken queries recorded under multiple environments~\cite{heigold2025massive}. While MSEB provides valuable coverage for evaluating general auditory representations, SVQ is built around simplified queries and acoustic conditions defined by recording environments rather than explicitly controlled and graded perturbations. Consequently, it does not provide an IR-aligned, end-to-end robustness benchmark for spoken query \emph{to text retrieval} under increasing acoustic stress. SQuTR fills this gap by extending established IR benchmarks to spoken queries with preserved corpora and relevance judgments, together with controlled, graded acoustic conditions.

\begin{figure*}[t]
    \centering
    \includegraphics[width=1.0\textwidth]{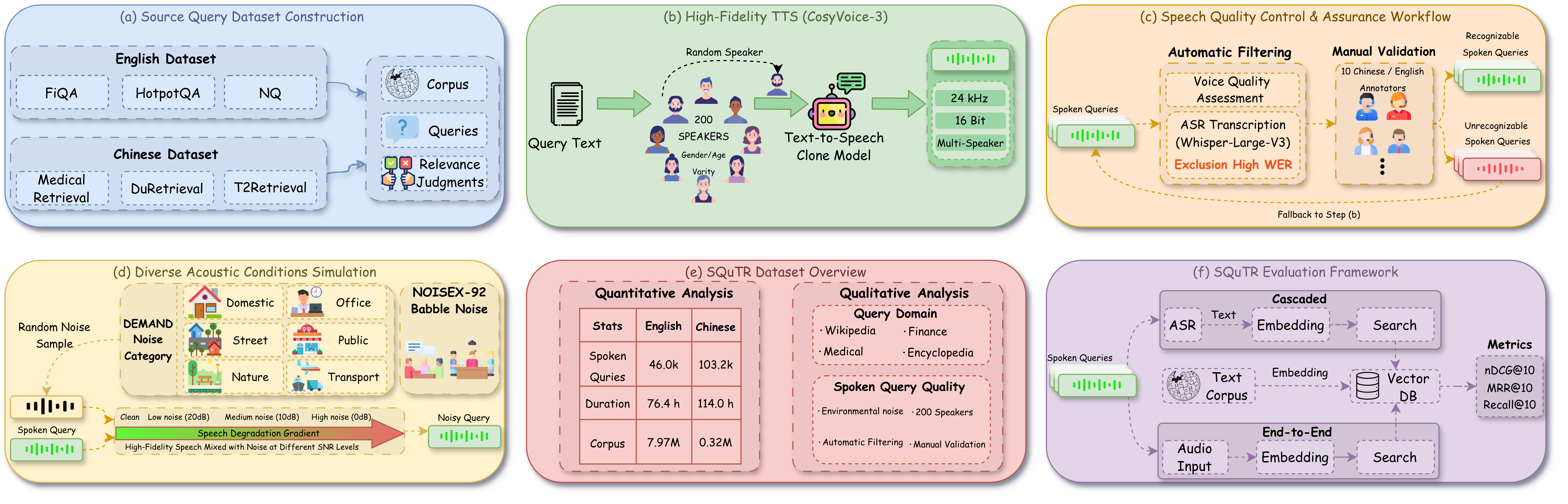}
    \caption{The SQuTR benchmark pipeline overview. (a-c) Construction of high-fidelity spoken queries from six IR benchmarks using rigorous quality control. (d) Simulation of four acoustic conditions under diverse environmental noise. (e-f) Dataset statistics and the unified evaluation framework for both cascaded and end-to-end retrieval systems.}
    \label{fig:pipeline}
\end{figure*}

\section{Dataset Construction}
SQuTR is designed to enable controlled and reproducible evaluation of spoken query to text retrieval under varying acoustic conditions.
An overview of the benchmark pipeline is provided in Figure~\ref{fig:pipeline}.

\subsection{Source Queries}
\label{sec:source}

\begin{table}[t]
\centering
\caption{Source datasets used in SQuTR. Query: \#queries; Corp.: corpus size; $Q_{\textit{len}}$/$D_{\textit{len}}$: avg. query/document length.}
\scriptsize
\resizebox{\columnwidth}{!}{%
\begin{tabular}{l l r r r r}
\toprule
\textbf{Dataset} & \textbf{Domain} & \textbf{Query} & \textbf{Corp.} & \textbf{$Q_{\textit{len}}$} & \textbf{$D_{\textit{len}}$} \\
\midrule
NQ~\cite{kwiatkowski2019natural}               & Wikipedia & 3{,}452  & 2.68M & 9.16  & 78.88 \\
HotpotQA~\cite{yang2018hotpotqa}         & Wikipedia & 7{,}405  & 5.23M & 17.61 & 46.30 \\
FiQA~\cite{de2018inf}             & Finance & 648      & 57.6K & 10.77 & 132.32 \\
DuRetrieval~\cite{qiu2022dureader-retrieval}      & Encyclopedia & 2{,}000  & 0.1M & 9.29  & 398.59 \\
MedicalRetrieval~\cite{long2022multicprmultidomainchinese} & Medical & 1{,}000  & 0.1M & 17.94 & 122.04 \\
T2Retrieval~\cite{xie2023t2ranking}      & Encyclopedia & 22{,}812 & 0.12M & 10.94 & 874.12 \\
\bottomrule
\end{tabular}%
}
\label{tab:squtr_datasets}
\end{table}

\begin{table}[t]
    \centering
    \caption{Summary of acoustic conditions used in SQuTR.}
    \label{tab:acoustic_conditions}
    \footnotesize
\begin{tabular}{l l c}
\toprule
\textbf{Condition} & \textbf{Description} & \textbf{SNR} \\
\midrule
Clean & Clean speech (no noise) & -- \\
Low Noise & Environmental noise recordings, high SNR & 20 dB \\
Medium Noise & Environmental noise recordings, moderate SNR & 10 dB \\
High Noise & Environmental noise recordings, low SNR & 0 dB \\
\bottomrule
\end{tabular}
\end{table}

SQuTR reuses queries from six widely used IR benchmarks in English and Chinese, as summarized in Table~\ref{tab:squtr_datasets}. 

For the English benchmarks (FiQA~\cite{de2018inf}, HotpotQA~\cite{yang2018hotpotqa}, and Natural Questions~\cite{kwiatkowski2019natural}), we use the test splits as queries, with document corpora and relevance judgments following MTEB~\cite{muennighoff2023mteb}. For the Chinese benchmarks (MedicalRetrieval~\cite{long2022multicprmultidomainchinese}, DuRetrieval~\cite{qiu2022dureader-retrieval}, and T2Retrieval~\cite{xie2023t2ranking}), we also use the test splits, following the configurations in C-MTEB~\cite{xiao2024c}. In all cases, document collections and relevance annotations are inherited unchanged from the original benchmarks.

We keep queries in their original form to preserve natural language characteristics and semantic complexity. The resulting set spans factoid QA, multi-hop QA, and general IR, and includes both short keyword-style queries and longer natural-language questions. Overall, SQuTR contains 37{,}317 unique queries.

\subsection{Speech Generation}
As illustrated in Figure~\ref{fig:pipeline}(b), we synthesize speech from text using CosyVoice-3~\cite{du2025cosyvoice3inthewildspeech} to construct spoken queries under controlled conditions. We fix key generation parameters (speaking rate, pitch, and random seeds) for reproducibility, and apply standard text normalization, including number verbalization, punctuation normalization, and consistent handling of English abbreviations.

We use voice profiles from 200 speakers with varying gender, age, and accents. Queries are grouped by language (English/Chinese), and speakers are uniformly sampled within each group. To mitigate synthesis artifacts, we generate three candidate renditions per query and retain the one with the lowest WER/CER under a reference ASR system.

Audio is generated at 24~kHz with 16-bit depth as clean base signals for subsequent noise injection. We perform automatic checks for synthesis failures and acoustic anomalies and regenerate affected samples when needed.

\subsection{Acoustic Conditions}

\label{sec:acoustic_conditions}
To systematically assess retrieval robustness under acoustic perturbations, we generate spoken queries under a set of controlled noise conditions starting from clean speech. SQuTR defines four acoustic conditions (Clean, Low Noise, Medium Noise, and High Noise), which differ only in noise intensity, as summarized in Table~\ref{tab:acoustic_conditions}.

Additive noise signals follow the DEMAND noise database~\cite{thiemann2013diverse} and cover a diverse set of environmental recordings, including public transportation, office spaces, household environments, and public venues. Babble noise from NOISEX-92~\cite{varga1993assessment} is additionally included. These noise sources are applied at predefined signal-to-noise ratio (SNR) levels to construct the Low, Medium, and High Noise conditions.
We further summarize the noise-type distribution in Section~\ref{sec:dataset_statistics} (Figure~\ref{fig:dataset_stats}).

Additive noise is applied using a global RMS-based scaling procedure to achieve
the target signal-to-noise ratio (SNR). Given clean speech $x[n]$ and a noise
signal $d[n]$, the noisy signal is constructed as
\begin{equation}
y[n] = x[n] + \alpha \cdot d[n]
\end{equation}
where the scaling factor $\alpha$ is computed based on the root mean square (RMS)
energy of the speech and noise signals:
\begin{equation}
\alpha = \frac{\mathrm{RMS}(x)}{\mathrm{RMS}(d)} \cdot 10^{-\frac{\mathrm{SNR}_{\mathrm{dB}}}{20}}
\end{equation}
RMS energy is computed after trimming leading and trailing silence to ensure
that the specified SNR reflects the active speech content. The resulting signal
is normalized to prevent clipping.

\subsection{Quality Control}
\label{sec:quality_control}
Figure~\ref{fig:pipeline}(c) outlines the quality control pipeline, which combines automated filtering with human verification.
Automated audio quality checks are used to detect synthesis artifacts, abnormal volume levels, and audio truncation. Clean speech samples are further transcribed using a high-performance ASR system based on Whisper-Large-v3~\cite{radford2023robust}, and WER is computed against the original text to identify samples that may require further inspection.


In the human verification stage, ten bilingual annotators perform auditory checks following standardized guidelines. Verification criteria include speech naturalness and clarity, correctness of the assigned noise condition, and semantic consistency with the original query. Samples that fail manual inspection are discarded and regenerated. All audio included in SQuTR undergoes at least one round of human verification.

\subsection{Dataset Statistics}
\label{sec:dataset_statistics}

\begin{table}[t]
\centering
\caption{Core statistics of SQuTR. Total speech duration sums over all four acoustic conditions.}
\label{tab:squtr_stats}
\setlength{\tabcolsep}{4pt}        
\renewcommand{\arraystretch}{1.0} 
\setlength{\aboverulesep}{0pt}     
\setlength{\belowrulesep}{0pt}
\small
\resizebox{0.9\columnwidth}{!}{%
\begin{tabular}{l r r r}
\toprule
\textbf{Metric} & \textbf{English} & \textbf{Chinese} & \textbf{Total} \\
\midrule
\#Unique Queries          & 11{,}505  & 25{,}812  & 37{,}317 \\
\#Speakers                & 100       & 100       & 200 \\
Average Query Length        & 13.48  & 11.08  & 11.82 \\
Total Speech Duration & 76.4 h   & 114.0 h   & 190.4 h \\
Average Speech Duration & 5.98 s   & 3.98 s   & 4.59 s \\
\#Evaluation instances    & 46{,}020 & 103{,}248 & 149{,}268 \\
\bottomrule
\end{tabular}%
}
\end{table}

The final SQuTR dataset contains \textbf{37,317 unique queries} spanning English and Chinese. In our benchmark, an evaluation instance is defined as a \emph{(query, acoustic condition)} pair. For each query, we generate spoken realizations under \textbf{four acoustic conditions}, resulting in \textbf{149,268 evaluation instances} across all settings. Table~\ref{tab:squtr_stats} summarizes the dataset, including the number of queries and speakers per language, total speech duration, and the overall scale of the benchmark. By combining realistic query distributions with systematic and controlled acoustic variation, SQuTR provides a solid foundation for robustness evaluation of spoken query to retrieval.

\begin{figure}[t]
    \centering
    \includegraphics[width=\linewidth]{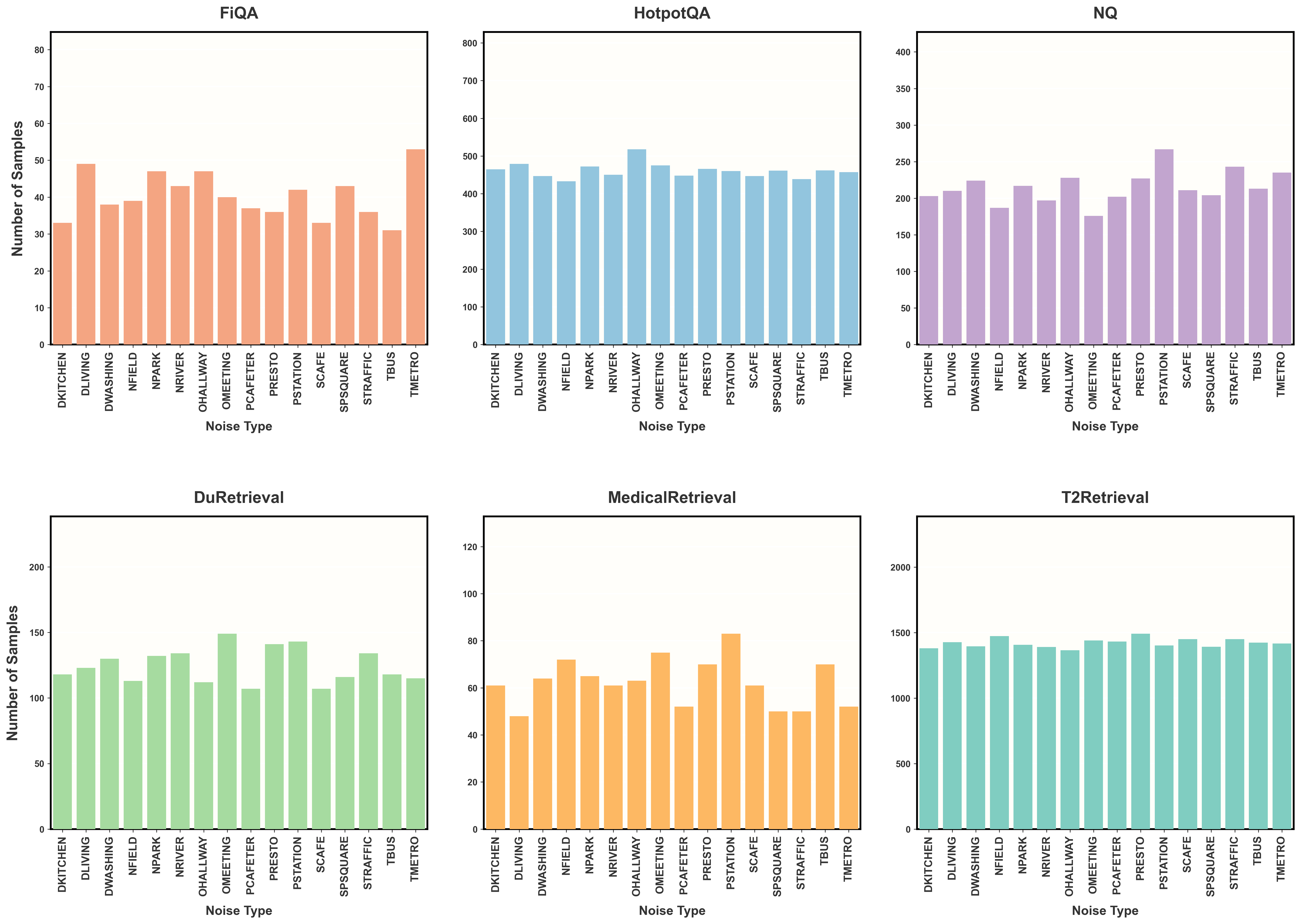}
    \vspace{-10pt}
    \caption{Distribution of noise categories in the SQuTR.}
    \label{fig:dataset_stats}
\end{figure}

Beyond scale, we further analyze the \emph{diversity} of acoustic environments represented in SQuTR. Figure~\ref{fig:dataset_stats} shows the distribution of noise categories, covering \textbf{16 distinct environmental classes}. By maintaining a balanced representation across these noise types, SQuTR enables evaluation across a wide spectrum of acoustic interference, without bias toward specific noise scenarios.

\subsection{Evaluation Protocol}
\label{subsec:evaluation_protocol}

As shown in Figure~\ref{fig:pipeline}(f), we define a unified evaluation protocol for spoken query retrieval, treating each system as a single pipeline that maps an input audio signal to a ranked list of text documents. In contrast to conventional practices that evaluate ASR and retrieval components separately, SQuTR evaluates retrieval performance directly from spoken input under different acoustic conditions. 

To assess performance, we utilize standard information retrieval metrics based on existing relevance judgments. We adopt nDCG@10 as the primary metric to evaluate both the presence and ranking position of relevant documents. Complementary metrics include Recall@k, which measures the coverage of relevant documents, and MRR@k, which assesses the capability to retrieve the first relevant result early. All metrics are computed on the final ranked list to reflect user-facing robustness against acoustic perturbations.

\section{Experiments}


\subsection{Experimental Setup}

\subsubsection{Datasets}
We construct SQuTR by reusing queries from six widely used text retrieval benchmarks~\cite{kwiatkowski2019natural, yang2018hotpotqa, long2022multicprmultidomainchinese, qiu2022dureader-retrieval, xie2023t2ranking, de2018inf}, including three English and three Chinese datasets, which are defined in Section~\ref{sec:source}.

\subsubsection{Acoustic Conditions}
We evaluate all systems on SQuTR under the four acoustic conditions defined in Section~\ref{sec:acoustic_conditions} and follow the evaluation protocol in
Section~\ref{subsec:evaluation_protocol}.
Specifically, we use noise samples corresponding to metro stations (\textit{TMETRO}), living rooms (\textit{DLIVING}), parks (\textit{PARK}), hallways (\textit{OHALLWAY}), riversides (\textit{NRIVER}), public squares (\textit{SPSQUARE}), train stations (\textit{PSTATION}), meeting rooms (\textit{OMEETING}), open fields (\textit{NFIELD}), washing environments (\textit{DWASHING}), cafeterias (\textit{PCAFETER}), restaurants (\textit{PRESTO}), street traffic (\textit{STRAFFIC}), cafés (\textit{SCAFE}), kitchens (\textit{DKITCHEN}), and buses (\textit{TBUS}). In addition, we include babble noise from the NOISEX-92 corpus (\textit{BABBLE}) to model background speech interference.

\begin{table}[t]
\centering
\caption{\textbf{ASR Performance.} Character Error Rate (CER) for Chinese (Paraformer-Large) and Word Error Rate (WER) for English (Whisper-Large-v3) under different noise levels.}
\label{tab:asr_performance}
\renewcommand{\arraystretch}{1.1}
\setlength{\tabcolsep}{8pt}
\begin{tabular}{lcc}
\toprule
\textbf{Noise Condition} & \textbf{Chinese (CER)} & \textbf{English (WER)} \\
\midrule
Clean & 2.71\% & 3.33\% \\
Low Noise (20dB) & 2.97\% & 4.10\% \\
Medium Noise (10dB) & 3.39\% & 4.48\% \\
High Noise (0dB) & 7.14\% & 7.75\% \\
\bottomrule
\end{tabular}
\vspace{-10pt}
\end{table}

\begin{table*}[t]
\centering

\caption{Retrieval performance on Chinese and English sub-datasets under varying acoustic conditions. (nDCG and MRR refer to nDCG@10 and MRR@10. Results for \textit{All-MiniLM-L6-v2} and \textit{Stella-EN-400M-v5} on Chinese datasets are marked with "--" because these models were not trained on Chinese text.)}
\label{tab:main_results_wide}
\renewcommand{\arraystretch}{1.2} 
\setlength{\tabcolsep}{2.5pt} 

\resizebox{\textwidth}{!}{%
\begin{tabular}{l cc cc cc cc cc cc cc cc cc cc}
\toprule 
\multirow{3}{*}{\textbf{Model Configuration}} & \multicolumn{10}{c}{\textbf{Chinese Sub-dataset}} & \multicolumn{10}{c}{\textbf{English Sub-dataset}} \\
\cmidrule(lr){2-11} \cmidrule(lr){12-21}
 & \multicolumn{2}{c}{\textbf{Text}} & \multicolumn{2}{c}{\textbf{Clean}} & \multicolumn{2}{c}{\textbf{Low (20dB)}} & \multicolumn{2}{c}{\textbf{Medium (10dB)}} & \multicolumn{2}{c}{\textbf{High (0dB)}} & \multicolumn{2}{c}{\textbf{Text}} & \multicolumn{2}{c}{\textbf{Clean}} & \multicolumn{2}{c}{\textbf{Low (20dB)}} & \multicolumn{2}{c}{\textbf{Medium (10dB)}} & \multicolumn{2}{c}{\textbf{High (0dB)}} \\
\cmidrule(lr){2-3} \cmidrule(lr){4-5} \cmidrule(lr){6-7} \cmidrule(lr){8-9} \cmidrule(lr){10-11} \cmidrule(lr){12-13} \cmidrule(lr){14-15} \cmidrule(lr){16-17} \cmidrule(lr){18-19} \cmidrule(lr){20-21}
 & \textbf{nDCG} & \textbf{MRR} & \textbf{nDCG} & \textbf{MRR} & \textbf{nDCG} & \textbf{MRR} & \textbf{nDCG} & \textbf{MRR} & \textbf{nDCG} & \textbf{MRR} & \textbf{nDCG} & \textbf{MRR} & \textbf{nDCG} & \textbf{MRR} & \textbf{nDCG} & \textbf{MRR} & \textbf{nDCG} & \textbf{MRR} & \textbf{nDCG} & \textbf{MRR} \\
\midrule
\multicolumn{21}{l}{\textbf{\textit{Cascaded Systems (ASR: Paraformer-Large [Zh] / Whisper-Large-v3 [En])}}} \\
\midrule
\multicolumn{21}{l}{\textit{\textbf{Lexical Baseline}}} \\
BM25 & 0.4843 & 0.5756 & 0.4380 & 0.5246 & 0.4366 & 0.5229 & 0.4313 & 0.5177 & 0.4061 & 0.4895 & 0.3912 & 0.4547 & 0.3586 & 0.4197 & 0.3570 & 0.4177 & 0.3555 & 0.4157 & 0.3374 & 0.3956 \\
\midrule
\multicolumn{21}{l}{\textit{\textbf{Dense (BGE Series)}}} \\
BGE-Small-\textit{(zh/en)}-v1.5 & 0.6871 & 0.7446 & 0.6491 & 0.7064 & 0.6454 & 0.7025 & 0.6402 & 0.6978 & 0.6025 & 0.6593 & 0.5345 & 0.5936 & 0.5070 & 0.5665 & 0.5035 & 0.5632 & 0.4992 & 0.5590 & 0.4756 & 0.5328 \\
BGE-Base-\textit{(zh/en)}-v1.5 & 0.7509 & 0.7890 & 0.7157 & 0.7557 & 0.7126 & 0.7523 & 0.7059 & 0.7457 & 0.6670 & 0.7075 & 0.5578 & 0.6130 & 0.5260 & 0.5845 & 0.5220 & 0.5801 & 0.5194 & 0.5766 & 0.4962 & 0.5513 \\
BGE-Large-\textit{(zh/en)}-v1.5 & 0.7662 & 0.8032 & 0.7306 & 0.7682 & 0.7274 & 0.7644 & 0.7212 & 0.7585 & 0.6801 & 0.7177 & 0.5801 & 0.6299 & 0.5521 & 0.6058 & 0.5493 & 0.6040 & 0.5466 & 0.6015 & 0.5194 & 0.5721 \\
BGE-M3-\textit{dense} & 0.7320 & 0.7756 & 0.6937 & 0.7381 & 0.6914 & 0.7359 & 0.6864 & 0.7315 & 0.6460 & 0.6912 & 0.5711 & 0.6368 & 0.5397 & 0.6035 & 0.5389 & 0.6034 & 0.5360 & 0.5996 & 0.5097 & 0.5686 \\
\midrule
\multicolumn{21}{l}{\textit{\textbf{Dense (Other)}}} \\
EmbeddingGemma-300M & 0.6952 & 0.7446 & 0.6626 & 0.7122 & 0.6603 & 0.7105 & 0.6554 & 0.7047 & 0.6188 & 0.6681 & 0.6029 & 0.6617 & 0.5797 & 0.6402 & 0.5775 & 0.6373 & 0.5747 & 0.6350 & 0.5497 & 0.6069 \\
Stella-EN-400M-v5 & -- & -- & -- & -- & -- & -- & -- & -- & -- & -- & 0.6198 & 0.6786 & 0.6017 & 0.6599 & 0.5986 & 0.6573 & 0.5962 & 0.6546 & 0.5706 & 0.6255 \\
All-MiniLM-L6-v2 & -- & -- & -- & -- & -- & -- & -- & -- & -- & -- & 0.4241 & 0.4862 & 0.3893 & 0.4433 & 0.3854 & 0.4387 & 0.3841 & 0.4368 & 0.3637 & 0.4136 \\
Multilingual-E5-Large & 0.7479 & 0.7900 & 0.7099 & 0.7528 & 0.7070 & 0.7503 & 0.7008 & 0.7447 & 0.6592 & 0.7019 & 0.5719 & 0.6323 & 0.5398 & 0.6006 & 0.5369 & 0.5967 & 0.5356 & 0.5960 & 0.5115 & 0.5698 \\
\midrule
\multicolumn{21}{l}{\textit{\textbf{Dense (Qwen3 Series)}}} \\
Qwen3-Embedding-0.6B & 0.7405 & 0.7840 & 0.7072 & 0.7512 & 0.7050 & 0.7492 & 0.6992 & 0.7438 & 0.6613 & 0.7063 & 0.5504 & 0.6234 & 0.5288 & 0.5988 & 0.5274 & 0.5978 & 0.5246 & 0.5961 & 0.5026 & 0.5697 \\
Qwen3-Embedding-4B & 0.7936 & 0.8237 & 0.7660 & 0.7978 & 0.7632 & 0.7958 & 0.7573 & 0.7899 & 0.7193 & 0.7528 & 0.6488 & 0.7110 & 0.6252 & 0.6886 & 0.6227 & 0.6860 & 0.6206 & 0.6839 & 0.5947 & 0.6565 \\
Qwen3-Embedding-8B & \textbf{0.8033} & \textbf{0.8315} & \textbf{0.7760} & \textbf{0.8057} & \textbf{0.7741} & \textbf{0.8041} & \textbf{0.7686} & \textbf{0.7988} & \textbf{0.7302} & \textbf{0.7608} & \textbf{0.6686} & \textbf{0.7253} & \textbf{0.6450} & \textbf{0.7041} & \textbf{0.6424} & \textbf{0.7021} & \textbf{0.6405} & \textbf{0.7000} & \textbf{0.6120} & \textbf{0.6690} \\
\midrule
\multicolumn{21}{l}{\textbf{\textit{End-to-End Systems (Direct Audio Input - No ASR)}}} \\
\midrule
Omni-Embed-Nemotron-3B & -- & -- & 0.6648 & 0.7201 & 0.6614 & 0.7179 & 0.6507 & 0.7067 & 0.5742 & 0.6314 & -- & -- & 0.5712 & 0.5394 & 0.5680 & 0.5369 & 0.5605 & 0.5289 & 0.5236 & 0.4959 \\
\bottomrule
\end{tabular}%
}
\vspace{-10pt}
\end{table*}



\subsubsection{Evaluated Systems}
\mbox{}\\
\textbf{Cascaded Systems (ASR + Retrieval)}:
Cascaded systems decompose spoken query retrieval into ASR followed by text-based retrieval. We pair high-performance ASR frontends with a diverse set of text retrieval backends. Specifically, we use \textbf{Whisper-Large-v3}~\cite{radford2023robust} for English and \textbf{Paraformer-Large}~\cite{gao2022paraformer} for Chinese.

On the retrieval side, we evaluate 12 backends covering both lexical and dense paradigms.  
(1) \textbf{Lexical baseline:} \textbf{BM25}~\cite{robertson2009probabilistic}. We implement BM25 using Anserini~\cite{lin2016toward}, with the default Lucene parameters ($k_1=0.9$, $b=0.4$). 
(2) \textbf{Dense retrievers:} We include models of varying scales and design choices, including the \textbf{BGE} series~\cite{xiao2024c, chen2024bge} (Small, Base, Large, and M3), \textbf{Qwen3-Embedding}~\cite{zhang2025qwen3} (0.6B, 4B, and 8B), and \textbf{Stella-v5-400M}~\cite{zhang2024jasper}. \emph{For BGE-M3, we report results using its dense-only variant for consistency with other dense retrievers.}
We further evaluate several widely used strong baselines, including \textbf{EmbeddingGemma-300M}~\cite{vera2025embeddinggemma}, \textbf{All-MiniLM-L6-v2}~\cite{wang2020minilm}, and \textbf{Multilingual-E5-Large-Instruct}~\cite{wang2024multilingual}.

\textbf{End-to-End Systems}: 
In contrast to cascaded pipelines, the end-to-end system directly maps spoken queries to retrieval representations without explicit transcription. We evaluate \textbf{Omni-Embed-Nemotron-3B}~\cite{xu2025omni}, which projects raw audio signals into a shared embedding space for retrieval.

\begin{figure}[t]
\centering
\includegraphics[width=0.9\linewidth]{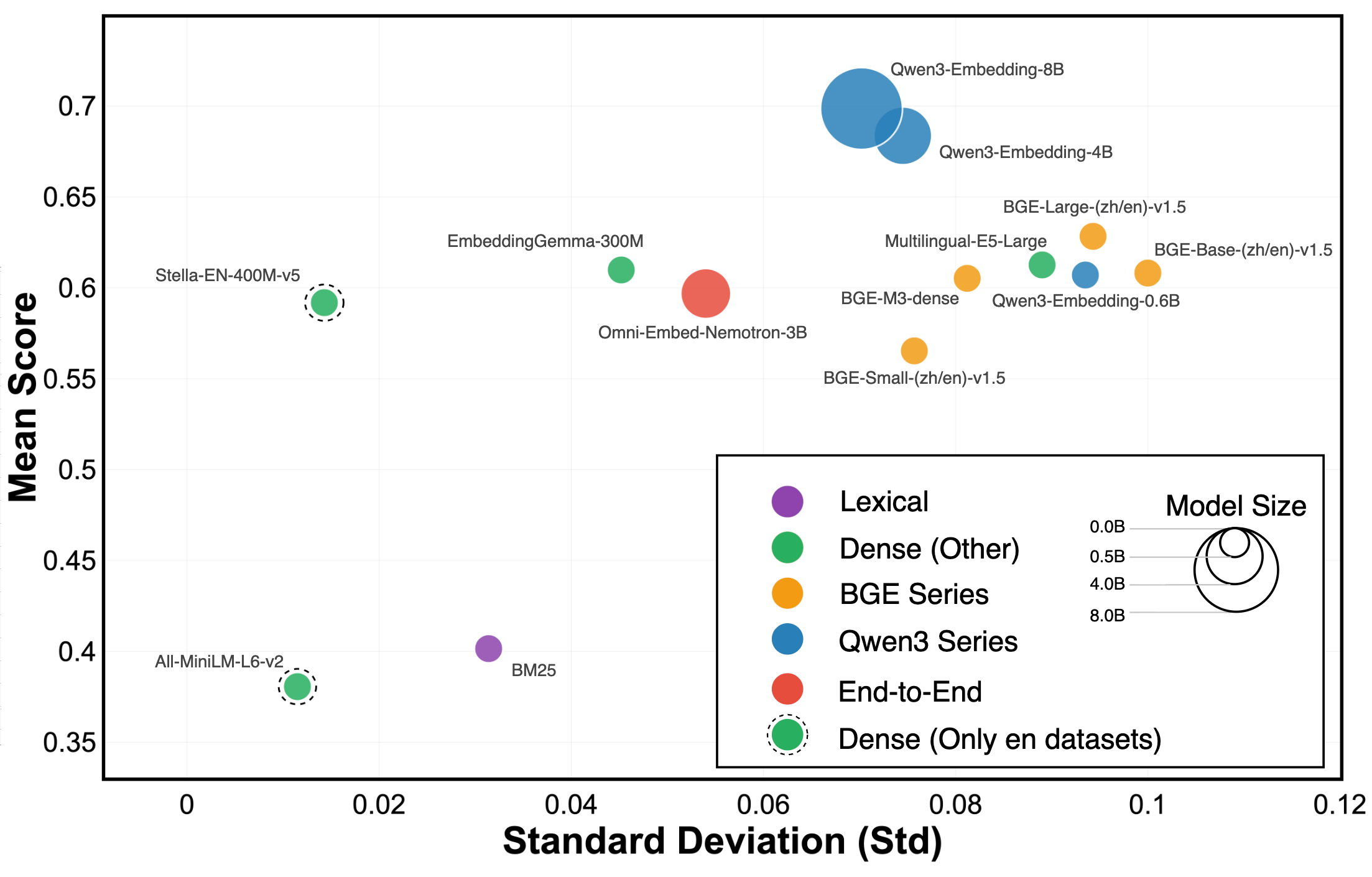}
\vspace{-10pt}
\caption{\textbf{Accuracy vs. Stability Trade-off.} The scatter plot positions systems based on Mean NDCG@10 (Y-axis) and Standard Deviation (X-axis). Models enclosed in \textbf{dashed lines} are evaluated on English datasets only.}
\vspace{-20pt}
\label{fig:stability_tradeoff}
\end{figure}

\begin{table*}[t]
\centering
\caption{Retrieval Performance across different ASR and Acoustic Conditions. nDCG and MRR refer
to nDCG@10 and MRR@10. }
\label{tab:ablation_asr_full_optimized}

\scriptsize
\renewcommand{\arraystretch}{0.95}
\setlength{\tabcolsep}{3pt}

\begin{tabular*}{\textwidth}{@{\extracolsep{\fill}} l l c cc cc c cc cc}
\toprule
& & \multicolumn{5}{c}{\textbf{Chinese Sub-dataset}} & \multicolumn{5}{c}{\textbf{English Sub-dataset}} \\
\cmidrule(lr){3-7} \cmidrule(lr){8-12}
\textbf{ASR Model} & \textbf{Noise Condition} &
\multirow{2}{*}{\textbf{CER(\%)}} &
\multicolumn{2}{c}{\textbf{BM25}} &
\multicolumn{2}{c}{\textbf{Qwen3-Embedding-8B}} &
\multirow{2}{*}{\textbf{WER(\%)}} &
\multicolumn{2}{c}{\textbf{BM25}} &
\multicolumn{2}{c}{\textbf{Qwen3-Embedding-8B}} \\
\cmidrule(lr){4-5} \cmidrule(lr){6-7} \cmidrule(lr){9-10} \cmidrule(lr){11-12}
& &
& \textbf{nDCG} & \textbf{MRR}
& \textbf{nDCG} & \textbf{MRR}
& 
& \textbf{nDCG} & \textbf{MRR}
& \textbf{nDCG} & \textbf{MRR} \\
\midrule

\multirow{4}{*}{SenseVoice-Small~\cite{an2024funaudiollm}}
& Clean               & 5.32  & 0.4090 & 0.4937 & 0.7562 & 0.7875 & 8.82  & 0.3759 & 0.3375 & 0.5991 & 0.6571 \\
& Low Noise (20dB)    & 6.22  & 0.4061 & 0.4915 & 0.7524 & 0.7839 & 9.46  & 0.3757 & 0.3363 & 0.5985 & 0.6561 \\
& Medium Noise (10dB) & 6.44  & 0.4013 & 0.4861 & 0.7476 & 0.7792 & 10.00 & 0.3699 & 0.3309 & 0.5941 & 0.6516 \\
& High Noise (0dB)    & 11.99 & 0.3734 & 0.4528 & 0.7109 & 0.7423 & 13.32 & 0.3436 & 0.3113 & 0.5599 & 0.6141 \\
\midrule

\multirow{4}{*}{Fun-ASR-Nano-2512~\cite{an2025fun}}
& Clean               & 3.08 & 0.4430 & 0.5298 & 0.7670 & 0.7963 & 6.47  & 0.3266 & 0.3837 & 0.6253 & 0.6837 \\
& Low Noise (20dB)    & 3.23 & 0.4419 & 0.5286 & 0.7675 & 0.7965 & 6.74  & 0.3253 & 0.3822 & 0.6215 & 0.6792 \\
& Medium Noise (10dB) & 3.47 & 0.4389 & 0.5259 & 0.7655 & 0.7948 & 6.93  & 0.3238 & 0.3809 & 0.6207 & 0.6784 \\
& High Noise (0dB)    & 6.72 & 0.4179 & 0.5017 & 0.7351 & 0.7644 & 10.03 & 0.3074 & 0.3626 & 0.5897 & 0.6464 \\
\midrule

\multirow{4}{*}{GLM-ASR-Nano-2512~\cite{zeng2025glm}}
& Clean               & 4.77  & 0.4305 & 0.5158 & 0.7526 & 0.7831 & 6.58 & 0.3256 & 0.3811 & 0.6322 & 0.6921 \\
& Low Noise (20dB)    & 4.73  & 0.4297 & 0.5152 & 0.7532 & 0.7839 & 6.37 & 0.3264 & 0.3818 & 0.6306 & 0.6897 \\
& Medium Noise (10dB) & 5.08  & 0.4256 & 0.5105 & 0.7483 & 0.7787 & 6.25 & 0.3260 & 0.3817 & 0.6284 & 0.6878 \\
& High Noise (0dB)    & 11.44 & 0.3905 & 0.4696 & 0.6941 & 0.7245 & 9.13 & 0.3094 & 0.3620 & 0.5970 & 0.6522 \\
\midrule

\multirow{4}{*}{Whisper-Large-V3~\cite{radford2023robust}}
& Clean               & 8.77  & 0.3718 & 0.4525 & 0.7019 & 0.7359 & 3.33 & 0.3586 & 0.4197 & 0.6450 & 0.7041 \\
& Low Noise (20dB)    & 8.96  & 0.3661 & 0.4471 & 0.6987 & 0.7334 & 4.10 & 0.3570 & 0.4177 & 0.6424 & 0.7021 \\
& Medium Noise (10dB) & 9.62  & 0.3589 & 0.4387 & 0.6929 & 0.7268 & 4.48 & 0.3555 & 0.4157 & 0.6405 & 0.7000 \\
& High Noise (0dB)    & 17.31 & 0.3020 & 0.3705 & 0.6128 & 0.6460 & 7.75 & 0.3374 & 0.3956 & 0.6120 & 0.6690 \\
\midrule

\multirow{4}{*}{Qwen3-ASR-1.7B~\cite{shi2026qwen3}}
& Clean               & 3.07 & 0.4416 & 0.5271 & 0.7655 & 0.7957 & 4.49 & 0.3503 & 0.4098 & 0.6407 & 0.6991 \\
& Low Noise (20dB)    & 3.19 & 0.4405 & 0.5255 & 0.7648 & 0.7955 & 4.90 & 0.3427 & 0.4015 & 0.6384 & 0.6977 \\
& Medium Noise (10dB) & 3.40 & 0.4368 & 0.5219 & 0.7611 & 0.7917 & 5.15 & 0.3403 & 0.3980 & 0.6357 & 0.6951 \\
& High Noise (0dB)    & 5.43 & 0.4198 & 0.5027 & 0.7351 & 0.7654 & 6.98 & 0.3290 & 0.3859 & 0.6150 & 0.6713 \\
\bottomrule
\end{tabular*}
\vspace{-10pt}
\end{table*}

\begin{table}[t]
\centering
\caption{Impact of Whisper ASR Model Size on Retrieval Performance: BM25 vs. Qwen3-Embedding-8B. nDCG and MRR refer to nDCG@10 and MRR@10.}
\vspace{-5pt}
\label{tab:whisper_ablation_full}

\newcolumntype{Y}{>{\centering\arraybackslash}p{1.6cm}}

\setlength{\tabcolsep}{3pt} 
\renewcommand{\arraystretch}{0.95}
\renewcommand{\multirowsetup}{\centering}

\resizebox{\columnwidth}{!}{
\begin{tabular}{l c l c Y Y Y Y}
\toprule
& & & & 
\multicolumn{2}{c}{\textbf{BM25}} & 
\multicolumn{2}{c}{\textbf{Qwen3-Embedding-8B}} \\

\cmidrule(lr){5-6} \cmidrule(lr){7-8}

\multirow{-2}{*}[0.5ex]{\textbf{ASR Model}} &
\multirow{-2}{*}[0.5ex]{\textbf{Params}} &
\multirow{-2}{*}[0.5ex]{\textbf{Noise Condition}} &
\multirow{-2}{*}[0.5ex]{\textbf{WER(\%)}} &
\textbf{nDCG } & \textbf{MRR} &
\textbf{nDCG} & \textbf{MRR} \\
\midrule

\multirow{4}{*}{Whisper-Tiny} & \multirow{4}{*}{39M}
& Clean               & 10.26 & 0.2966 & 0.3511 & 0.5862 & 0.6455 \\
& & Low Noise (20dB)    & 10.57 & 0.2963 & 0.3506 & 0.5828 & 0.6403 \\
& & Medium Noise (10dB) & 12.46 & 0.2852 & 0.3387 & 0.5665 & 0.6245 \\
& & High Noise (0dB)    & 26.48 & 0.2344 & 0.2808 & 0.4666 & 0.5191 \\
\midrule

\multirow{4}{*}{Whisper-Base} & \multirow{4}{*}{74M}
& Clean               & 8.12  & 0.3147 & 0.3719 & 0.6022 & 0.6612 \\
& & Low Noise (20dB)    & 8.30  & 0.3140 & 0.3706 & 0.5996 & 0.6579 \\
& & Medium Noise (10dB) & 9.34  & 0.3088 & 0.3644 & 0.5915 & 0.6498 \\
& & High Noise (0dB)    & 18.19 & 0.2691 & 0.3206 & 0.5148 & 0.5689 \\
\midrule

\multirow{4}{*}{Whisper-Small} & \multirow{4}{*}{244M}
& Clean               & 6.17  & 0.3358 & 0.3944 & 0.6246 & 0.6848 \\
& & Low Noise (20dB)    & 6.18  & 0.3365 & 0.3955 & 0.6229 & 0.6820 \\
& & Medium Noise (10dB) & 6.78  & 0.3339 & 0.3927 & 0.6190 & 0.6775 \\
& & High Noise (0dB)    & 12.01 & 0.3050 & 0.3599 & 0.5681 & 0.6239 \\
\midrule

\multirow{4}{*}{Whisper-Medium} & \multirow{4}{*}{769M}
& Clean               & 5.21  & 0.3465 & 0.4061 & 0.6322 & 0.6920 \\
& & Low Noise (20dB)    & 5.11  & 0.3476 & 0.4078 & 0.6329 & 0.6923 \\
& & Medium Noise (10dB) & 5.45  & 0.3451 & 0.4046 & 0.6294 & 0.6878 \\
& & High Noise (0dB)    & 9.72  & 0.3231 & 0.3807 & 0.5931 & 0.6496 \\
\midrule

\multirow{4}{*}{Whisper-Large-v3} & \multirow{4}{*}{1550M}
& Clean               & 3.33  & 0.3586 & 0.4197 & 0.6450 & 0.7041 \\
& & Low Noise (20dB)    & 4.10  & 0.3570 & 0.4177 & 0.6424 & 0.7021 \\
& & Medium Noise (10dB) & 4.48  & 0.3555 & 0.4157 & 0.6405 & 0.7000 \\
& & High Noise (0dB)    & 7.75  & 0.3374 & 0.3956 & 0.6120 & 0.6690 \\
\bottomrule
\end{tabular}
}
\vspace{-15pt}
\end{table}

\subsection{Main Results}
Table~\ref{tab:main_results_wide} validates SQuTR's effectiveness in quantifying robustness and discriminating between architectural capabilities under graded acoustic noise.
As noise increases (Clean $\to$ Low $\to$ High), retrieval performance degrades across systems, but the degradation patterns differ by model scale and design: traditional dense retrievers (e.g., BGE-Large) show a more pronounced monotonic drop, whereas large generative models remain comparatively stable (e.g., Qwen3-Embedding-8B sustains stronger performance at 0dB), indicating that SQuTR provides sufficient discriminative power for robustness evaluation.

We also observe that spoken-query retrieval underperforms the text upper bound even in clean conditions, and the gap widens with noise. For example, Qwen3-Embedding-8B (Chinese) drops from 0.8033 on \textit{Text} to 0.7760 on \textit{Clean Speech}. This gap is consistent with error propagation in cascaded systems (ASR errors affecting downstream retrieval) and the inherent difficulty of aligning speech and text representations in end-to-end models.
Finally, SQuTR remains a challenging testbed. Even the strongest model, Qwen3-Embedding-8B, does not match text performance under noise (0.7302 at \textit{High Noise} vs.\ 0.8033 on \textit{Text}), and end-to-end models still lag behind strong cascaded baselines, leaving substantial headroom for improving noise-robust spoken query retrieval.

\subsection{Robustness and Stability Analysis}
We examine the trade-off between accuracy (\textbf{mean nDCG@10}) and stability (\textbf{standard deviation}, $\sigma$), where $\sigma$ is computed over all language–condition combinations (two languages $\times$ four acoustic conditions). Figure~\ref{fig:stability_tradeoff} visualizes this distribution. All-MiniLM-L6-v2 and Stella-EN-400M-v5 (dashed circles) are evaluated on English only and excluded from the main comparison.

First, there is a clear stability gap between sparse and dense architectures. BM25 remains comparatively stable ($\sigma=0.031$), while BERT-based dense retrievers (e.g., BGE-Base, $\sigma=0.100$) exhibit substantially higher variance, indicating greater sensitivity to acoustic perturbations despite strong clean-condition performance. Second, model scaling and direct modality modeling improve robustness. Within the Qwen3 series, increasing model size from 0.6B to 8B reduces variance ($\sigma: 0.094 \rightarrow 0.070$) while improving mean effectiveness. The end-to-end model Omni-Embed-Nemotron-3B ($\sigma=0.054$) further narrows this gap, suggesting that larger capacity and direct speech-to-embedding mapping mitigate instability.

\subsection{Ablation Study}
\label{subsec:ablation}

Table~\ref{tab:whisper_ablation_full} confirms that scaling ASR models (e.g., Whisper-Tiny to Whisper-Large-v3) improves retrieval metrics. However, the choice of retrieval backend proves more critical than ASR size: notably, the smallest Whisper-Tiny using dense retrieval (Qwen3-Embedding-8B) significantly outperforms the largest Whisper-Large-v3 using BM25. This highlights that semantic robustness effectively compensates for ASR transcription errors where lexical matching fails.
Despite the performance gap, lexical and dense retrieval exhibit highly consistent degradation trends, maintaining stability from \textit{Clean} to \textit{Medium Noise} (10dB) before dropping sharply at \textit{High Noise} (0dB).
We attribute the minor BM25 performance fluctuations under \textit{Low Noise} (20dB) to stochastic variance rather than noise benefits. Since modern ASR models saturate at 20dB (treating it effectively as clean), random variations in error patterns (e.g., hallucinations vs. deletions) can occasionally favor exact keyword matching. In contrast, dense retrieval filters these lexical instabilities, resulting in a smoother, monotonic decline.

\section{Conclusion}
We introduce SQuTR, a controllable and reproducible benchmark for evaluating spoken query to text retrieval under graded acoustic noise. The benchmark comprises a large-scale spoken-query dataset derived from widely used text retrieval benchmarks and augmented with systematically controlled acoustic conditions.
SQuTR provides a unified framework to assess robustness across lexical, dense, and end-to-end retrieval paradigms.
Our experiments reveal consistent performance degradation as noise increases, and highlight architectural and scaling differences in robustness.
These findings suggest that noise-robust spoken retrieval remains an open challenge. We hope SQuTR will facilitate more systematic and comparable research on robustness in spoken query to text retrieval.



\bibliographystyle{ACM-Reference-Format}

\clearpage

\appendix

\section{Mathematical Formulation of Acoustic Simulation}
\label{sec:app_acoustic_math}

To ensure the reproducibility of the SQuTR acoustic environment, we formally define the signal degradation process based on global energy statistics. The distorted signal $y[n]$ is modeled as a linear superposition of the reverberant speech and additive noise:

\begin{equation}
    y[n] = (x * h)[n] + \alpha \cdot d[n]
\end{equation}

where $n$ is the discrete time index. $x[n]$ represents the clean speech synthesized by CosyVoice-3. 
Prior to processing, all audio components are strictly resampled to a unified rate of 24kHz to prevent spectral aliasing.

The convolution $(x * h)[n]$ incorporates room acoustics via the Room Impulse Response (RIR) $h[n]$. We define the target signal $s[n] = (x * h)[n]$ as the holistic reverberant speech.
To ensure the Signal-to-Noise Ratio (SNR) accurately reflects the active speech level, standard leading and trailing silence trimming is applied to $x[n]$ before energy calculation.

The additive noise sequence $d[n]$ is randomly sampled from the DEMAND/NOISEX-92 datasets. The scaling factor $\alpha$ is derived from the global Root Mean Square (RMS) amplitude:

\begin{equation}
    \text{RMS}(s) = \sqrt{\frac{1}{N}\sum_{n=0}^{N-1} |s[n]|^2}, \quad 
    \text{RMS}(d) = \sqrt{\frac{1}{N}\sum_{n=0}^{N-1} |d[n]|^2}
\end{equation}

where $N$ denotes the signal length. Given a target $\text{SNR}_{\text{dB}}$, the mixing coefficient $\alpha$ is solved as:

\begin{equation}
    \alpha = \frac{\text{RMS}(s)}{\text{RMS}(d)} \cdot 10^{-\frac{\text{SNR}_{\text{dB}}}{20}}
\end{equation}

Finally, the mixed signal $y[n]$ is globally normalized to fit within the standard PCM dynamic range, utilizing a headroom factor $\beta=0.9$ to prevent digital clipping.

\section{Automated Quality Control Protocol}
\label{sec:app_qc_algo}

To ensure the semantic integrity of the dataset, we define a rigorous filtering function $\Phi: \mathcal{X} \to \{0, 1\}$ that maps a generated audio sample $x$ to a binary acceptance decision. The process involves a cascade of objective metrics and subjective proxy evaluations.

The filtering pipeline, formally described in Algorithm 1 (Figure \ref{fig:qc_protocol}), integrates two distinct verification stages:
\begin{enumerate}
    \item \textbf{Objective Consistency Check:} We utilize a large-scale ASR model ($\mathcal{M}_{ASR}$) to verify if the synthesized speech $a_i$ aligns with the ground-truth text $t_i$. To mitigate minor formatting discrepancies, a normalization function $\mathcal{N}(\cdot)$ (lowercasing, punctuation removal) is applied before computing the Word Error Rate (WER).
    \item \textbf{Subjective Proxy Evaluation:} We employ a "Judge-LLM" paradigm utilizing two distinct Large Language Models, $\mathcal{M}_{Gemini}$ and $\mathcal{M}_{GPT}$, as independent evaluators. They score the audio quality $Q \in [1, 5]$ based on a Chain-of-Thought reasoning process.
\end{enumerate}

\begin{figure}[h]
    \centering
    \setlength{\fboxsep}{10pt} 
    \setlength{\fboxrule}{0.8pt} 
    
    \fcolorbox{black}{ricewhite}{%
        \begin{minipage}{0.9\linewidth}
            \small 
            \textbf{Algorithm 1:} Robust Multi-Stage Quality Filtering
            \par\noindent\hrulefill\par
            
            \smallskip\noindent\textbf{Input:} ASR Model $\mathcal{M}_{ASR}$ (Whisper-v3 for En, Paraformer for Zh)
            
            \textbf{Hyperparameters:}
            \begin{itemize}
                \item WER Threshold $\tau_{wer} = 0.30$
                \item Quality Score Threshold $\tau_{qual} = 4$
            \end{itemize}
            
            \textbf{Output:} Clean Dataset $\mathcal{D}_{clean}$
            
            \vspace{0.5em}
            \hrule
            \vspace{0.5em}
            
            \begin{enumerate}
                \item $\mathcal{D}_{clean} \leftarrow \emptyset$
                \item \textbf{for} each pair $(a_i, t_i) \in \mathcal{D}$ \textbf{do}
                
                \item \quad \textit{\textbf{// Stage 1: Objective ASR Verification}}
                \item \quad $\hat{t}_i \leftarrow \mathcal{M}_{ASR}(a_i)$ \hfill $\triangleright$ \textit{Transcribe Audio}
                \item \quad $\hat{t}'_i, t'_i \leftarrow \mathcal{N}(\hat{t}_i), \mathcal{N}(t_i)$ \hfill $\triangleright$ \textit{Text Normalization}
                \item \quad $\delta_{wer} \leftarrow \text{CalcWER}(\hat{t}'_i, t'_i)$
                
                \item \quad \textbf{if} $\delta_{wer} > \tau_{wer}$ \textbf{then}
                \item \quad \quad \textbf{continue} \hfill $\triangleright$ \textit{Reject: High mismatch}
                
                \item \quad \textit{\textbf{// Stage 2: Dual-LLM Consensus}}
                \item \quad $S_{scores} \leftarrow \emptyset$
                \item \quad \textbf{for} $\mathcal{M}_k \in \mathcal{E}$ \textbf{do} \hfill $\triangleright$ \textit{Parallel Evaluation}
                \item \quad \quad $s_k, \text{reason}_k \leftarrow \mathcal{M}_k.\text{Eval}(a_i, t_i | \text{Prompt}_{CoT})$
                \item \quad \quad $S_{scores} \leftarrow S_{scores} \cup \{s_k\}$
                
                \item \quad \textbf{if} $\min(S_{scores}) \ge \tau_{qual}$ \textbf{then}
                \item \quad \quad $\mathcal{D}_{clean} \leftarrow \mathcal{D}_{clean} \cup \{(a_i, t_i)\}$ \hfill $\triangleright$ \textit{Accept}
                \item \quad \textbf{else}
                \item \quad \quad Discard $a_i$ (Optionally flag for manual review)
                
                \item \textbf{end for}
                \item \textbf{return} $\mathcal{D}_{clean}$
            \end{enumerate}
        \end{minipage}
    }
    \caption{Pseudocode for the automated quality control protocol. The pipeline enforces both semantic accuracy (via ASR) and perceptual fidelity (via LLM consensus).}
    \label{fig:qc_protocol}
\end{figure}
\section{Detailed Experimental Setup}
\label{sec:app_exp_setup}

To ensure full reproducibility, we strictly controlled the computational environment, inference hyperparameters, and model-specific configurations.

\subsection{Computational Environment}
All experiments were conducted on a high-performance computing cluster. The detailed hardware and software specifications are listed in Table \ref{tab:env_config}.

\begin{table}[h]
    \centering
    \scriptsize
    \caption{Global hardware and software configuration.}
    \label{tab:env_config}
    \renewcommand{\arraystretch}{1.2}
    \begin{tabular}{p{0.15\linewidth} p{0.25\linewidth} p{0.5\linewidth}}
        \hline\hline
        \textbf{Type} & \textbf{Component} & \textbf{Specification} \\
        \hline
        \textbf{H/W} & GPU & NVIDIA A100-SXM4 (80GB) \\
        \hline
        \textbf{S/W} & OS & Ubuntu 24.04 LTS \\  
             & Framework & PyTorch 2.5.1 \\ 
             & Transformers & v4.57.1 \\
        \hline
        \textbf{ASR} & Decoding & Greedy \\
                     & Precision & FP16 \\
        \hline\hline
    \end{tabular}
\end{table}

\subsection{ASR Inference Protocol}
\label{sec:app_asr_protocol}

To comprehensively evaluate the impact of ASR robustness on retrieval, we employed a multi-model strategy covering different parameter scales and languages.

\paragraph{\textbf{Input Preprocessing}}
Although the SQuTR dataset is mastered at a sampling rate of 24kHz to preserve high-frequency acoustic details (as detailed in Appendix~\ref{sec:app_acoustic_math}), most pre-trained ASR systems are optimized for 16kHz input. To prevent spectral aliasing and timescale modification artifacts during inference, we applied \textbf{on-the-fly downsampling} to 16kHz for all ASR models using \texttt{torchaudio} (Lanczos resampling kernel).

\paragraph{\textbf{Model Configurations}}
We evaluated the following systems:

\begin{itemize}
    \item \textbf{English (Whisper Family):} We utilized OpenAI's \texttt{whisper} series:
    \begin{itemize}
        \item \textit{Tiny} (39M): \texttt{openai/whisper-tiny}
        \item \textit{Base} (74M): \texttt{openai/whisper-base}
        \item \textit{Small} (244M): \texttt{openai/whisper-small}
        \item \textit{Medium} (769M): \texttt{openai/whisper-medium}
        
    \end{itemize}
    
    \item \textbf{Chinese:} We adopted Paraformer.
    \begin{itemize}
        \item \textit{Paraformer-Large}: \texttt{funasr/paraformer-large}
    \end{itemize}
    
    \item \textbf{Multilingual Baselines:} To assess cross-lingual robustness, we also included:
    \begin{itemize}
        \item \textit{SenseVoice-Small}: \texttt{FunAudioLLM/SenseVoiceSmall}
        \item \textit{Whisper-Large-v3}: \texttt{openai/whisper-large-v3}
        \item \textit{GLM-ASR-Nano-2512}: \texttt{zai-org/GLM-ASR-Nano-2512}
        \item \textit{Fun-ASR-Nano-2512}: \texttt{FunAudioLLM/Fun-ASR-Nano-2512}
        \item \textit{Qwen3-ASR-1.7B}: \texttt{Qwen/Qwen3-ASR-1.7B}
    \end{itemize}
\end{itemize}

For all ASR inference, we used deterministic greedy decoding with a forced language token to prevent identification errors.

\subsection{Retrieval Model Configuration}
We evaluated 15 distinct cascaded retrieval architectures. Modern embedding models often require specific **task instructions** (prefixes) to activate their asymmetric retrieval capabilities. We strictly adhered to the official prompt templates as detailed in Table \ref{tab:model_prompts}.

\begin{table}[h]
    \centering
    \scriptsize
    \caption{Model-specific query instructions (Prefixes) for all embedding models.}
    \label{tab:model_prompts}
    \renewcommand{\arraystretch}{1.3}
    \begin{tabular}{p{0.3\linewidth} p{0.6\linewidth}}
        \hline\hline
        \textbf{Model Family} & \textbf{Query Instruction / Prompt} \\
        \hline
        \textbf{Lexical} & \\
        BM25 & N/A ($k_1=0.9, b=0.4$) \\
        \hline
        \textbf{BGE Series v1.5} & "Represent this sentence for searching relevant passages:" \\
        \hline
        \textbf{BGE-M3} & N/A (Dense vector-only mode) \\
        \hline
        \textbf{Multilingual-E5-Large} & "query: "\\
        \hline
        \textbf{EmbeddingGemma-300M} & "task: search result | query: " \\
        \hline
        \textbf{All-MiniLM-L6-v2} & N/A \\
        \hline
        \textbf{Stella-EN-400M-v5} & "Instruct: Given a web search query, retrieve relevant passages that answer the query.
        Query:"\\
        \hline
        \textbf{Qwen3 Embedding Series} & "Given a web search query, retrieve relevant passages that answer the query"\\
        \hline
        \textbf{Omni-Embed-Nemotron-3B} & N/A \\
        \hline\hline
    \end{tabular}
\end{table}

\subsection{Evaluation Metrics Standard}
We report the normalized Discounted Cumulative Gain at rank 10 (nDCG@10). The implementation strictly follows the NIST \texttt{trec\_eval} standard.
\begin{itemize}
    \item \textbf{Relevance Mapping:} For datasets with graded relevance (e.g., FiQA), we preserve original grades. For binary datasets (e.g., NQ), relevance is mapped to $\{0, 1\}$.
    \item \textbf{Global Ranking:} For multi-GPU inference, embeddings were gathered across devices to ensure accurate global ranking statistics.
\end{itemize}

\begin{figure}[htbp]
    \centering
    \setlength{\fboxsep}{8pt} 
    \setlength{\fboxrule}{0.8pt}
    
    \fcolorbox{black}{verylightgray}{
        \begin{minipage}{0.90\linewidth}
            \footnotesize 
            
            \textbf{\texttt{[SYSTEM INSTRUCTION]}}
            \par\noindent\hrulefill\par
            
            \textbf{Role:} You are an expert \textbf{Audio Forensic Analyst} and \textbf{Linguistic Quality Specialist}.
            
            \textbf{Task:} You will be provided with a \texttt{Reference Text} and an \texttt{Audio Clip}. Your goal is to evaluate the audio's fidelity and quality with extreme rigor.
            
            \vspace{0.3em}
            \textbf{Evaluation Protocol (Step-by-Step):}
            \begin{enumerate}
                \setlength\itemsep{0em} 
                \item \textbf{Listen \& Transcribe:} Mentally transcribe the audio. Compare it strictly against the Reference Text.
                \item \textbf{Check for Fatal Errors:} 
                \begin{itemize}
                    \item \textit{Truncation:} Is the start or end of the sentence cut off?
                    \item \textit{Hallucination:} Does the audio contain words not in the text?
                    \item \textit{Omission:} Are any words skipped?
                \end{itemize}
                \item \textbf{Assess Signal Quality:} Detect background static, metallic robotic artifacts, or unnatural pitch shifts.
            \end{enumerate}
            
            \vspace{0.3em}
            \textbf{Scoring Rubric (Strict):}
            \begin{itemize}
                \setlength\itemsep{0em}
                \item \textbf{5.0 (Perfect):} Indistinguishable from human recording. 100\% text match. Zero noise.
                \item \textbf{4.0 (Good):} Clear and correct. Slight digital "flavor" but fully intelligible.
                \item \textbf{3.0 (Acceptable):} Noticeable noise or slight mispronunciation, but meaning is preserved.
                \item \textbf{2.0 (Bad):} Hard to understand. Significant artifacts. Missing 1-2 non-critical words.
                \item \textbf{1.0 (Critical Failure):} Wrong text, severe truncation, or pure noise.
            \end{itemize}
            
            \vspace{0.3em}
            \textbf{Constraint Checklist:}
            \begin{itemize}
                \setlength\itemsep{0em}
                \item Do NOT be lenient. Be critical of minor glitches.
                \item If the audio is truncated (incomplete sentence), the max score is 2.0.
                \item Output MUST be valid JSON.
            \end{itemize}

            \vspace{0.5em}
            \textbf{\texttt{[USER INPUT]}}
            \par\noindent\hrulefill\par
            \textbf{Reference Text:} "\{TRANSCRIPT\}" \\
            \textbf{Audio Data:} <Audio\_Attachment>
            
            \vspace{0.5em}
            \textbf{\texttt{[MODEL OUTPUT SCHEMA]}}
            \par\noindent\hrulefill\par
            
            \begin{tabular}{@{}l}
                \texttt{\{} \\
                \texttt{~~"analysis\_trace": \{} \\
                \texttt{~~~~"text\_alignment": "Exact Match / Mismatch",} \\
                \texttt{~~~~"detected\_artifacts": ["None", "Static", "Truncation"],} \\
                \texttt{~~~~"prosody\_assessment": "Natural / Robotic / Monotone"} \\
                \texttt{~~\},} \\
                \texttt{~~"fatal\_error\_flag": false,} \\
                \texttt{~~"reasoning\_summary": "The audio is clear...",} \\
                \texttt{~~"final\_quality\_score": <Float 1.0-5.0>} \\
                \texttt{\}}
            \end{tabular}
            
        \end{minipage}
    }
    \caption{The advanced Chain-of-Thought (CoT) prompt template used for automated quality control. The prompt includes explicit failure mode definitions and specific handling for TTS edge cases (e.g., truncation).}
    \label{fig:advanced_prompt}
\end{figure}

\section{Robust Evaluation Prompts}
\label{sec:app_prompts}

To ensure the validity and consistency of our "LLM-as-a-Judge" pipeline, we developed a sophisticated prompt template incorporating \textbf{Role-Playing}, \textbf{Chain-of-Thought (CoT)} reasoning, and \textbf{Negative Constraints}.

The prompt explicitly enforces an "Audio Forensic Analyst" persona to minimize hallucination. The JSON output schema is strictly defined to ensure parsing reliability. The complete configuration is presented in Figure \ref{fig:advanced_prompt}.

\section{Extended Experimental Results}
\label{sec:app_extended_results}

We present the full retrieval diagnostics across all six sub-datasets in Tables~\ref{tab:fiqa_main_results} through~\ref{tab:t2retrieval_asr_merged}.

\newcommand{\resrow}[1]{
    \multirow{3}{*}{#1} 
    & nDCG@10 & 0.0001 & 0.0001 & 0.0001 & 0.0001 & 0.0001 \\
    & MRR@10  & 0.0001 & 0.0001 & 0.0001 & 0.0001 & 0.0001 \\
    & R@10    & 0.0001 & 0.0001 & 0.0001 & 0.0001 & 0.0001 \\
    \cmidrule{1-7}
}

\newcommand{\tablebodycontent}{
    \multicolumn{7}{l}{\textit{\textbf{I. Lexical \& Lightweight Baselines}}} \\
    \cmidrule{1-7}
    \resrow{BM25}
    \resrow{MiniLM-L6}
    
    \multicolumn{7}{l}{\textit{\textbf{II. Dense Retrieval (BGE Series v1.5)}}} \\
    \cmidrule{1-7}
    \resrow{bge-small-en}
    \resrow{bge-base-en}
    \resrow{bge-large-en}
    \resrow{bge-m3 (Hybrid)}
    
    \multicolumn{7}{l}{\textit{\textbf{III. Multilingual \& Mid-sized}}} \\
    \cmidrule{1-7}
    \resrow{Multi-E5-Large}
    \resrow{Stella-400M-v5}
    \resrow{Stella-1.5B-v5}
    \resrow{Gemma-300M}
    
    \multicolumn{7}{l}{\textit{\textbf{IV. Large Language Model Embeddings}}} \\
    \cmidrule{1-7}
    \resrow{Qwen2-1.5B}
    \resrow{Qwen2-7B-Inst}
    \resrow{Llama-3-8B}
    \resrow{NV-Embed-v1}
    \resrow{SFR-Mistral}
}

\begin{table*}[htbp]
\centering
\caption{Main experimental results on the FiQA dataset under different acoustic conditions. The metrics nDCG, MRR, and Recall denote nDCG@10, MRR@10, and Recall@10, respectively. Best results are highlighted in bold.}
\label{tab:fiqa_main_results}
\resizebox{\textwidth}{!}{%
\begin{tabular}{l ccc ccc ccc ccc ccc}
\toprule
\multicolumn{16}{l}{\textbf{Dataset: FiQA}} \\
\midrule
\multirow{2}{*}{\textbf{Model Config}} & \multicolumn{3}{c}{\textbf{Text}} & \multicolumn{3}{c}{\textbf{Clean}} & \multicolumn{3}{c}{\textbf{Low (20dB)}} & \multicolumn{3}{c}{\textbf{Medium (10dB)}} & \multicolumn{3}{c}{\textbf{High (0dB)}} \\
\cmidrule(lr){2-4} \cmidrule(lr){5-7} \cmidrule(lr){8-10} \cmidrule(lr){11-13} \cmidrule(lr){14-16}
 & nDCG & MRR & Recall & nDCG & MRR & Recall & nDCG & MRR & Recall & nDCG & MRR & Recall & nDCG & MRR & Recall \\
\midrule
BM25 & 0.2388 & 0.3002 & 0.2995 & 0.2335 & 0.2946 & 0.2869 & 0.2329 & 0.2943 & 0.2887 & 0.2333 & 0.2934 & 0.2903 & 0.2214 & 0.2787 & 0.2736 \\
BGE-Small-(zh/en)-v1.5 & 0.4030 & 0.4879 & 0.4640 & 0.3962 & 0.4767 & 0.4635 & 0.3926 & 0.4745 & 0.4568 & 0.3895 & 0.4714 & 0.4525 & 0.3750 & 0.4519 & 0.4390 \\
BGE-Base-(zh/en)-v1.5 & 0.4062 & 0.4859 & 0.4804 & 0.3933 & 0.4763 & 0.4643 & 0.3883 & 0.4707 & 0.4569 & 0.3911 & 0.4720 & 0.4620 & 0.3766 & 0.4521 & 0.4485 \\
BGE-Large-(zh/en)-v1.5 & 0.4487 & 0.5327 & 0.5130 & 0.4333 & 0.5158 & 0.4987 & 0.4331 & 0.5195 & 0.4976 & 0.4332 & 0.5203 & 0.4983 & 0.4130 & 0.4951 & 0.4775 \\
BGE-M3-dense & 0.4126 & 0.5046 & 0.4720 & 0.3948 & 0.4824 & 0.4574 & 0.3972 & 0.4860 & 0.4592 & 0.3940 & 0.4805 & 0.4540 & 0.3760 & 0.4531 & 0.4438 \\
EmbeddingGemma-300M & 0.4739 & 0.5467 & 0.5533 & 0.4666 & 0.5462 & 0.5421 & 0.4634 & 0.5398 & 0.5405 & 0.4631 & 0.5413 & 0.5399 & 0.4485 & 0.5210 & 0.5239 \\
Stella-EN-400M-v5 & 0.5494 & 0.6290 & 0.6225 & 0.5427 & 0.6225 & 0.6162 & 0.5392 & 0.6195 & 0.6121 & 0.5396 & 0.6186 & 0.6120 & 0.5231 & 0.5942 & 0.5977 \\
All-MiniLM-L6-v2 & 0.3687 & 0.4446 & 0.4414 & 0.3475 & 0.4195 & 0.4200 & 0.3419 & 0.4133 & 0.4122 & 0.3468 & 0.4176 & 0.4193 & 0.3306 & 0.3990 & 0.4036 \\
Multilingual-E5-Large & 0.4617 & 0.5439 & 0.5324 & 0.4391 & 0.5181 & 0.5150 & 0.4346 & 0.5120 & 0.5113 & 0.4366 & 0.5138 & 0.5147 & 0.4229 & 0.4983 & 0.4981 \\
Qwen3-Embedding-0.6B & 0.4701 & 0.5546 & 0.5472 & 0.4567 & 0.5376 & 0.5324 & 0.4561 & 0.5388 & 0.5332 & 0.4543 & 0.5397 & 0.5279 & 0.4406 & 0.5199 & 0.5195 \\
Qwen3-Embedding-4B & 0.5889 & 0.6692 & 0.6694 & 0.5763 & 0.6586 & 0.6542 & 0.5730 & 0.6547 & 0.6559 & 0.5738 & 0.6554 & 0.6537 & 0.5523 & 0.6293 & 0.6289 \\
\textbf{Qwen3-Embedding-8B} & \textbf{0.6162} & \textbf{0.6845} & \textbf{0.6990} & \textbf{0.5968} & \textbf{0.6653} & \textbf{0.6793} & \textbf{0.5949} & \textbf{0.6647} & \textbf{0.6786} & \textbf{0.5974} & \textbf{0.6678} & \textbf{0.6788} & \textbf{0.5723} & \textbf{0.6357} & \textbf{0.6571} \\
Omni-Embed-Nemotron-3B & -- & -- & -- & 0.4568 & 0.2922 & 0.5047 & 0.4548 & 0.2935 & 0.5003 & 0.4515 & 0.2894 & 0.4988 & 0.4299 & 0.2794 & 0.4738 \\
\bottomrule
\end{tabular}%
}
\end{table*}

\begin{table*}[htbp]
\centering
\caption{Main experimental results on the HotpotQA dataset under different acoustic conditions. The metrics nDCG, MRR, and Recall denote nDCG@10, MRR@10, and Recall@10, respectively. Best results are highlighted in bold.}
\label{tab:hotpotqa_main_results}
\resizebox{\textwidth}{!}{%
\begin{tabular}{l ccc ccc ccc ccc ccc}
\toprule
\multicolumn{16}{l}{\textbf{Dataset: HotpotQA}} \\
\midrule
\multirow{2}{*}{\textbf{Model Config}} & \multicolumn{3}{c}{\textbf{Text}} & \multicolumn{3}{c}{\textbf{Clean}} & \multicolumn{3}{c}{\textbf{Low (20dB)}} & \multicolumn{3}{c}{\textbf{Medium (10dB)}} & \multicolumn{3}{c}{\textbf{High (0dB)}} \\
\cmidrule(lr){2-4} \cmidrule(lr){5-7} \cmidrule(lr){8-10} \cmidrule(lr){11-13} \cmidrule(lr){14-16}
 & nDCG & MRR & Recall & nDCG & MRR & Recall & nDCG & MRR & Recall & nDCG & MRR & Recall & nDCG & MRR & Recall \\
\midrule
BM25 & 0.6291 & 0.8004 & 0.6531 & 0.5464 & 0.7098 & 0.5727 & 0.5426 & 0.7040 & 0.5697 & 0.5399 & 0.7008 & 0.5667 & 0.5128 & 0.6678 & 0.5394 \\
BGE-Small-(zh/en)-v1.5 & 0.6993 & 0.8413 & 0.7279 & 0.6197 & 0.7649 & 0.6519 & 0.6158 & 0.7608 & 0.6487 & 0.6119 & 0.7563 & 0.6451 & 0.5801 & 0.7211 & 0.6128 \\
BGE-Base-(zh/en)-v1.5 & 0.7259 & 0.8610 & 0.7573 & 0.6543 & 0.7953 & 0.6911 & 0.6492 & 0.7904 & 0.6868 & 0.6452 & 0.7846 & 0.6825 & 0.6119 & 0.7498 & 0.6466 \\
BGE-Large-(zh/en)-v1.5 & 0.7411 & 0.8613 & \textbf{0.7800} & 0.6708 & 0.8029 & 0.7106 & 0.6649 & 0.7955 & 0.7071 & 0.6621 & 0.7922 & 0.7041 & 0.6273 & 0.7541 & 0.6695 \\
BGE-M3-dense & 0.6944 & 0.8466 & 0.7182 & 0.6294 & 0.7822 & 0.6597 & 0.6287 & 0.7824 & 0.6580 & 0.6264 & 0.7797 & 0.6565 & 0.5941 & 0.7412 & 0.6245 \\
EmbeddingGemma-300M & 0.7008 & 0.8581 & 0.7290 & 0.6574 & 0.8166 & 0.6891 & 0.6556 & 0.8147 & 0.6862 & 0.6525 & 0.8110 & 0.6835 & 0.6216 & 0.7760 & 0.6518 \\
Stella-EN-400M-v5 & 0.6874 & 0.8387 & 0.7204 & 0.6479 & 0.7972 & 0.6848 & 0.6448 & 0.7952 & 0.6808 & 0.6417 & 0.7917 & 0.6771 & 0.6100 & 0.7553 & 0.6464 \\
All-MiniLM-L6-v2 & 0.4650 & 0.6282 & 0.4867 & 0.4122 & 0.5560 & 0.4393 & 0.4077 & 0.5501 & 0.4344 & 0.4031 & 0.5437 & 0.4303 & 0.3797 & 0.5114 & 0.4069 \\
Multilingual-E5-Large & 0.6828 & 0.8298 & 0.7167 & 0.6165 & 0.7656 & 0.6530 & 0.6140 & 0.7622 & 0.6513 & 0.6118 & 0.7607 & 0.6479 & 0.5803 & 0.7232 & 0.6165 \\
Qwen3-Embedding-0.6B & 0.6506 & 0.8361 & 0.6600 & 0.5986 & 0.7786 & 0.6151 & 0.5962 & 0.7762 & 0.6132 & 0.5943 & 0.7737 & 0.6109 & 0.5652 & 0.7372 & 0.5835 \\
Qwen3-Embedding-4B & 0.7249 & 0.8831 & 0.7442 & 0.6791 & 0.8404 & 0.7047 & 0.6768 & 0.8383 & 0.7029 & 0.6728 & 0.8338 & 0.6991 & 0.6427 & 0.8013 & 0.6689 \\
\textbf{Qwen3-Embedding-8B} & \textbf{0.7424} & \textbf{0.8954} & 0.7637 & \textbf{0.6968} & \textbf{0.8550} & \textbf{0.7235} & \textbf{0.6931} & \textbf{0.8514} & \textbf{0.7200} & \textbf{0.6891} & \textbf{0.8458} & \textbf{0.7167} & \textbf{0.6574} & \textbf{0.8120} & \textbf{0.6850} \\
Omni-Embed-Nemotron-3B & -- & -- & -- & 0.6188 & 0.7351 & 0.6657 & 0.6139 & 0.7284 & 0.6629 & 0.6035 & 0.7175 & 0.6531 & 0.5590 & 0.6723 & 0.6057 \\
\bottomrule
\end{tabular}%
}
\end{table*}

\begin{table*}[htbp]
\centering
\caption{Main experimental results on the NQ dataset under different acoustic conditions. The metrics nDCG, MRR, and Recall denote nDCG@10, MRR@10, and Recall@10, respectively. Best results are highlighted in bold.}
\label{tab:nq_main_results}
\resizebox{\textwidth}{!}{%
\begin{tabular}{l ccc ccc ccc ccc ccc}
\toprule
\multicolumn{16}{l}{\textbf{Dataset: NQ}} \\
\midrule
\multirow{2}{*}{\textbf{Model Config}} & \multicolumn{3}{c}{\textbf{Text}} & \multicolumn{3}{c}{\textbf{Clean}} & \multicolumn{3}{c}{\textbf{Low (20dB)}} & \multicolumn{3}{c}{\textbf{Medium (10dB)}} & \multicolumn{3}{c}{\textbf{High (0dB)}} \\
\cmidrule(lr){2-4} \cmidrule(lr){5-7} \cmidrule(lr){8-10} \cmidrule(lr){11-13} \cmidrule(lr){14-16}
 & nDCG & MRR & Recall & nDCG & MRR & Recall & nDCG & MRR & Recall & nDCG & MRR & Recall & nDCG & MRR & Recall \\
\midrule
BM25 & 0.3056 & 0.2636 & 0.4731 & 0.2959 & 0.2547 & 0.4602 & 0.2954 & 0.2549 & 0.4579 & 0.2934 & 0.2529 & 0.4551 & 0.2780 & 0.2403 & 0.4291 \\
BGE-Small-(zh/en)-v1.5 & 0.5012 & 0.4516 & 0.7080 & 0.5052 & 0.4580 & 0.7029 & 0.5021 & 0.4542 & 0.6996 & 0.4962 & 0.4494 & 0.6895 & 0.4718 & 0.4254 & 0.6619 \\
BGE-Base-(zh/en)-v1.5 & 0.5412 & 0.4920 & 0.7464 & 0.5305 & 0.4819 & 0.7326 & 0.5285 & 0.4792 & 0.7294 & 0.5219 & 0.4731 & 0.7220 & 0.5002 & 0.4521 & 0.6964 \\
BGE-Large-(zh/en)-v1.5 & 0.5504 & 0.4957 & 0.7650 & 0.5522 & 0.4985 & 0.7629 & 0.5499 & 0.4969 & 0.7602 & 0.5446 & 0.4920 & 0.7536 & 0.5178 & 0.4671 & 0.7199 \\
BGE-M3-dense & 0.6063 & 0.5593 & 0.7942 & 0.5950 & 0.5459 & 0.7895 & 0.5909 & 0.5417 & 0.7862 & 0.5876 & 0.5387 & 0.7819 & 0.5590 & 0.5115 & 0.7469 \\
EmbeddingGemma-300M & 0.6342 & 0.5804 & 0.8447 & 0.6150 & 0.5580 & 0.8322 & 0.6136 & 0.5574 & 0.8287 & 0.6085 & 0.5526 & 0.8227 & 0.5789 & 0.5237 & 0.7908 \\
Stella-EN-400M-v5 & 0.6227 & 0.5683 & 0.8408 & 0.6143 & 0.5598 & 0.8322 & 0.6117 & 0.5572 & 0.8294 & 0.6074 & 0.5536 & 0.8233 & 0.5787 & 0.5269 & 0.7882 \\
All-MiniLM-L6-v2 & 0.4386 & 0.3859 & 0.6474 & 0.4081 & 0.3544 & 0.6181 & 0.4067 & 0.3528 & 0.6153 & 0.4023 & 0.3492 & 0.6097 & 0.3809 & 0.3304 & 0.5797 \\
Multilingual-E5-Large & 0.5712 & 0.5231 & 0.7675 & 0.5639 & 0.5181 & 0.7565 & 0.5622 & 0.5159 & 0.7574 & 0.5585 & 0.5135 & 0.7506 & 0.5314 & 0.4878 & 0.7183 \\
Qwen3-Embedding-0.6B & 0.5305 & 0.4795 & 0.7371 & 0.5310 & 0.4801 & 0.7385 & 0.5300 & 0.4784 & 0.7387 & 0.5253 & 0.4748 & 0.7305 & 0.5020 & 0.4520 & 0.7022 \\
Qwen3-Embedding-4B & 0.6327 & 0.5806 & 0.8388 & 0.6203 & 0.5669 & 0.8284 & 0.6183 & 0.5649 & 0.8262 & 0.6152 & 0.5625 & 0.8220 & 0.5891 & 0.5389 & 0.7880 \\
\textbf{Qwen3-Embedding-8B} & \textbf{0.6472} & \textbf{0.5961} & \textbf{0.8517} & \textbf{0.6415} & \textbf{0.5920} & \textbf{0.8430} & \textbf{0.6391} & \textbf{0.5901} & \textbf{0.8395} & \textbf{0.6350} & \textbf{0.5865} & \textbf{0.8342} & \textbf{0.6064} & \textbf{0.5592} & \textbf{0.8011} \\
Omni-Embed-Nemotron-3B & -- & -- & -- & 0.6380 & 0.5910 & 0.8302 & 0.6353 & 0.5888 & 0.8238 & 0.6264 & 0.5797 & 0.8160 & 0.5818 & 0.5360 & 0.7678 \\
\bottomrule
\end{tabular}%
}
\end{table*}

\begin{table*}[htbp]
\centering
\caption{Main experimental results on the DuRetrieval dataset under different acoustic conditions. The metrics nDCG, MRR, and Recall denote nDCG@10, MRR@10, and Recall@10, respectively. Best results are highlighted in bold.}
\label{tab:duretrieval_main_results}
\resizebox{\textwidth}{!}{%
\begin{tabular}{l ccc ccc ccc ccc ccc}
\toprule
\multicolumn{16}{l}{\textbf{Dataset: DuRetrieval}} \\
\midrule
\multirow{2}{*}{\textbf{Model Config}} & \multicolumn{3}{c}{\textbf{Text}} & \multicolumn{3}{c}{\textbf{Clean}} & \multicolumn{3}{c}{\textbf{Low (20dB)}} & \multicolumn{3}{c}{\textbf{Medium (10dB)}} & \multicolumn{3}{c}{\textbf{High (0dB)}} \\
\cmidrule(lr){2-4} \cmidrule(lr){5-7} \cmidrule(lr){8-10} \cmidrule(lr){11-13} \cmidrule(lr){14-16}
 & nDCG & MRR & Recall & nDCG & MRR & Recall & nDCG & MRR & Recall & nDCG & MRR & Recall & nDCG & MRR & Recall \\
\midrule
BM25 & 0.5613 & 0.7000 & 0.5714 & 0.5109 & 0.6404 & 0.5257 & 0.5093 & 0.6386 & 0.5241 & 0.5012 & 0.6309 & 0.5153 & 0.4706 & 0.5952 & 0.4842 \\
BGE-Small-(zh/en)-v1.5 & 0.7900 & 0.8776 & 0.8051 & 0.7551 & 0.8409 & 0.7753 & 0.7497 & 0.8369 & 0.7697 & 0.7444 & 0.8325 & 0.7653 & 0.6980 & 0.7831 & 0.7206 \\
BGE-Base-(zh/en)-v1.5 & 0.8507 & 0.9102 & 0.8728 & 0.8198 & 0.8818 & 0.8429 & 0.8135 & 0.8755 & 0.8366 & 0.8062 & 0.8674 & 0.8313 & 0.7570 & 0.8180 & 0.7818 \\
BGE-Large-(zh/en)-v1.5 & 0.8634 & 0.9201 & 0.8838 & 0.8348 & 0.8929 & 0.8581 & 0.8286 & 0.8867 & 0.8521 & 0.8227 & 0.8809 & 0.8466 & 0.7706 & 0.8272 & 0.7969 \\
BGE-M3-dense & 0.8397 & 0.9081 & 0.8626 & 0.8143 & 0.8844 & 0.8385 & 0.8106 & 0.8811 & 0.8352 & 0.8031 & 0.8746 & 0.8279 & 0.7534 & 0.8251 & 0.7793 \\
EmbeddingGemma-300M & 0.7879 & 0.8664 & 0.8131 & 0.7624 & 0.8423 & 0.7876 & 0.7574 & 0.8387 & 0.7827 & 0.7505 & 0.8292 & 0.7775 & 0.7086 & 0.7865 & 0.7358 \\
Stella-EN-400M-v5 & -- & -- & -- & -- & -- & -- & -- & -- & -- & -- & -- & -- & -- & -- & -- \\
All-MiniLM-L6-v2 & -- & -- & -- & -- & -- & -- & -- & -- & -- & -- & -- & -- & -- & -- & -- \\
Multilingual-E5-Large & 0.8487 & 0.9128 & 0.8663 & 0.8190 & 0.8852 & 0.8408 & 0.8137 & 0.8813 & 0.8349 & 0.8071 & 0.8755 & 0.8292 & 0.7554 & 0.8223 & 0.7790 \\
Qwen3-Embedding-0.6B & 0.8265 & 0.8965 & 0.8478 & 0.8021 & 0.8747 & 0.8245 & 0.7991 & 0.8725 & 0.8215 & 0.7911 & 0.8652 & 0.8137 & 0.7460 & 0.8188 & 0.7683 \\
Qwen3-Embedding-4B & 0.8884 & 0.9412 & 0.9020 & 0.8658 & 0.9223 & 0.8800 & 0.8615 & 0.9196 & 0.8755 & 0.8545 & 0.9129 & 0.8686 & 0.8121 & 0.8710 & 0.8270 \\
\textbf{Qwen3-Embedding-8B} & \textbf{0.8979} & \textbf{0.9477} & \textbf{0.9076} & \textbf{0.8768} & \textbf{0.9307} & \textbf{0.8868} & \textbf{0.8733} & \textbf{0.9273} & \textbf{0.8849} & \textbf{0.8670} & \textbf{0.9219} & \textbf{0.8785} & \textbf{0.8219} & \textbf{0.8781} & \textbf{0.8353} \\
Omni-Embed-Nemotron-3B & -- & -- & -- & 0.7612 & 0.8464 & 0.7889 & 0.7546 & 0.8420 & 0.7825 & 0.7470 & 0.8344 & 0.7764 & 0.6624 & 0.7514 & 0.6906 \\
\bottomrule
\end{tabular}%
}
\end{table*}

\begin{table*}[htbp]
\centering
\caption{Main experimental results on the MedicalRetrieval dataset under different acoustic conditions. The metrics nDCG, MRR, and Recall denote nDCG@10, MRR@10, and Recall@10, respectively. Best results are highlighted in bold.}
\label{tab:medical_retrieval_main_results}
\resizebox{\textwidth}{!}{%
\begin{tabular}{l ccc ccc ccc ccc ccc}
\toprule
\multicolumn{16}{l}{\textbf{Dataset: MedicalRetrieval}} \\
\midrule
\multirow{2}{*}{\textbf{Model Config}} & \multicolumn{3}{c}{\textbf{Text}} & \multicolumn{3}{c}{\textbf{Clean}} & \multicolumn{3}{c}{\textbf{Low (20dB)}} & \multicolumn{3}{c}{\textbf{Medium (10dB)}} & \multicolumn{3}{c}{\textbf{High (0dB)}} \\
\cmidrule(lr){2-4} \cmidrule(lr){5-7} \cmidrule(lr){8-10} \cmidrule(lr){11-13} \cmidrule(lr){14-16}
 & nDCG & MRR & Recall & nDCG & MRR & Recall & nDCG & MRR & Recall & nDCG & MRR & Recall & nDCG & MRR & Recall \\
\midrule
BM25 & 0.3110 & 0.2937 & 0.3670 & 0.2796 & 0.2625 & 0.3350 & 0.2796 & 0.2624 & 0.3350 & 0.2744 & 0.2575 & 0.3290 & 0.2570 & 0.2406 & 0.2240 \\
BGE-Small-(zh/en)-v1.5 & 0.4993 & 0.4738 & 0.5850 & 0.4707 & 0.4453 & 0.5560 & 0.4674 & 0.4401 & 0.5560 & 0.4612 & 0.4344 & 0.5480 & 0.4294 & 0.4037 & 0.5130 \\
BGE-Base-(zh/en)-v1.5 & 0.5649 & 0.5355 & 0.6600 & 0.5358 & 0.5063 & 0.6330 & 0.5353 & 0.5052 & 0.6330 & 0.5271 & 0.4980 & 0.6210 & 0.4948 & 0.4678 & 0.5830 \\
BGE-Large-(zh/en)-v1.5 & 0.5953 & 0.5692 & 0.6810 & 0.5629 & 0.5345 & 0.6570 & 0.5619 & 0.5320 & 0.6580 & 0.5546 & 0.5249 & 0.6500 & 0.5186 & 0.4912 & 0.6060 \\
BGE-M3-dense & 0.5424 & 0.5158 & 0.6280 & 0.4985 & 0.4703 & 0.5910 & 0.4980 & 0.4701 & 0.5900 & 0.4948 & 0.4669 & 0.5860 & 0.4604 & 0.4326 & 0.5530 \\
EmbeddingGemma-300M & 0.4990 & 0.4723 & 0.5840 & 0.4666 & 0.4389 & 0.5550 & 0.4669 & 0.4397 & 0.5540 & 0.4634 & 0.4359 & 0.5510 & 0.4324 & 0.4073 & 0.5140 \\
Stella-EN-400M-v5 & -- & -- & -- & -- & -- & -- & -- & -- & -- & -- & -- & -- & -- & -- & -- \\
All-MiniLM-L6-v2 & -- & -- & -- & -- & -- & -- & -- & -- & -- & -- & -- & -- & -- & -- & -- \\
Multilingual-E5-Large & 0.5609 & 0.5382 & 0.6430 & 0.5198 & 0.4947 & 0.6140 & 0.5186 & 0.4932 & 0.6130 & 0.5107 & 0.4862 & 0.6030 & 0.4753 & 0.4494 & 0.5700 \\
Qwen3-Embedding-0.6B & 0.5618 & 0.5356 & 0.6460 & 0.5251 & 0.4951 & 0.6210 & 0.5239 & 0.4939 & 0.6200 & 0.5182 & 0.4885 & 0.6130 & 0.4857 & 0.4584 & 0.5730 \\
Qwen3-Embedding-4B & 0.6207 & 0.5904 & 0.7180 & 0.5959 & 0.5643 & 0.6980 & 0.5943 & 0.5633 & 0.6940 & 0.5878 & 0.5564 & 0.6890 & 0.5514 & 0.5208 & 0.6500 \\
\textbf{Qwen3-Embedding-8B} & \textbf{0.6340} & \textbf{0.6043} & \textbf{0.7310} & \textbf{0.6090} & \textbf{0.5765} & \textbf{0.7160} & \textbf{0.6092} & \textbf{0.5774} & \textbf{0.7140} & \textbf{0.6034} & \textbf{0.5707} & \textbf{0.7110} & \textbf{0.5688} & \textbf{0.5349} & \textbf{0.6790} \\
Omni-Embed-Nemotron-3B & -- & -- & -- & 0.4930 & 0.4676 & 0.5740 & 0.4946 & 0.4695 & 0.5740 & 0.4808 & 0.4529 & 0.5700 & 0.4127 & 0.3876 & 0.4930 \\
\bottomrule
\end{tabular}%
}
\end{table*}

\begin{table*}[htbp]
\centering
\caption{Main experimental results on the T2Retrieval dataset under different acoustic conditions. The metrics nDCG, MRR, and Recall denote nDCG@10, MRR@10, and Recall@10, respectively. Best results are highlighted in bold.}
\label{tab:t2retrieval_main_results}
\resizebox{\textwidth}{!}{%
\begin{tabular}{l ccc ccc ccc ccc ccc}
\toprule
\multicolumn{16}{l}{\textbf{Dataset: T2Retrieval}} \\
\midrule
\multirow{2}{*}{\textbf{Model Config}} & \multicolumn{3}{c}{\textbf{Text}} & \multicolumn{3}{c}{\textbf{Clean}} & \multicolumn{3}{c}{\textbf{Low (20dB)}} & \multicolumn{3}{c}{\textbf{Medium (10dB)}} & \multicolumn{3}{c}{\textbf{High (0dB)}} \\
\cmidrule(lr){2-4} \cmidrule(lr){5-7} \cmidrule(lr){8-10} \cmidrule(lr){11-13} \cmidrule(lr){14-16}
 & nDCG & MRR & Recall & nDCG & MRR & Recall & nDCG & MRR & Recall & nDCG & MRR & Recall & nDCG & MRR & Recall \\
\midrule
BM25 & 0.5806 & 0.7330 & 0.5690 & 0.5236 & 0.6708 & 0.5157 & 0.5208 & 0.6677 & 0.5130 & 0.5183 & 0.6648 & 0.5105 & 0.4908 & 0.6327 & 0.4842 \\
BGE-Small-(zh/en)-v1.5 & 0.7721 & 0.8824 & 0.7591 & 0.7216 & 0.8331 & 0.7128 & 0.7192 & 0.8305 & 0.7106 & 0.7149 & 0.8264 & 0.7060 & 0.6802 & 0.7911 & 0.6730 \\
BGE-Base-(zh/en)-v1.5 & 0.8371 & 0.9214 & 0.8237 & 0.7914 & 0.8790 & 0.7817 & 0.7891 & 0.8762 & 0.7796 & 0.7843 & 0.8716 & 0.7753 & 0.7493 & 0.8368 & 0.7421 \\
BGE-Large-(zh/en)-v1.5 & 0.8399 & 0.9204 & 0.8295 & 0.7942 & 0.8771 & 0.7881 & 0.7917 & 0.8745 & 0.7858 & 0.7864 & 0.8696 & 0.7807 & 0.7511 & 0.8347 & 0.7467 \\
BGE-M3-dense & 0.8138 & 0.9029 & 0.8050 & 0.7684 & 0.8595 & 0.7633 & 0.7655 & 0.8564 & 0.7603 & 0.7614 & 0.8529 & 0.7562 & 0.7243 & 0.8160 & 0.7203 \\
EmbeddingGemma-300M & 0.7987 & 0.8951 & 0.7879 & 0.7588 & 0.8555 & 0.7517 & 0.7565 & 0.8531 & 0.7501 & 0.7523 & 0.8490 & 0.7457 & 0.7155 & 0.8105 & 0.7101 \\
Stella-EN-400M-v5 & -- & -- & -- & -- & -- & -- & -- & -- & -- & -- & -- & -- & -- & -- & -- \\
All-MiniLM-L6-v2 & -- & -- & -- & -- & -- & -- & -- & -- & -- & -- & -- & -- & -- & -- & -- \\
Multilingual-E5-Large & 0.8341 & 0.9189 & 0.8212 & 0.7909 & 0.8786 & 0.7813 & 0.7887 & 0.8763 & 0.7791 & 0.7845 & 0.8725 & 0.7752 & 0.7468 & 0.8340 & 0.7398 \\
Qwen3-Embedding-0.6B & 0.8333 & 0.9199 & 0.8185 & 0.7943 & 0.8838 & 0.7829 & 0.7920 & 0.8812 & 0.7813 & 0.7882 & 0.8776 & 0.7773 & 0.7522 & 0.8416 & 0.7435 \\
Qwen3-Embedding-4B & 0.8718 & 0.9394 & 0.8595 & 0.8362 & 0.9067 & 0.8278 & 0.8338 & 0.9045 & 0.8256 & 0.8296 & 0.9003 & 0.8218 & 0.7943 & 0.8665 & 0.7874 \\
\textbf{Qwen3-Embedding-8B} & \textbf{0.8780} & \textbf{0.9424} & \textbf{0.8663} & \textbf{0.8423} & \textbf{0.9098} & \textbf{0.8345} & \textbf{0.8398} & \textbf{0.9075} & \textbf{0.8319} & \textbf{0.8355} & \textbf{0.9037} & \textbf{0.8278} & \textbf{0.8000} & \textbf{0.8694} & \textbf{0.7944} \\
Omni-Embed-Nemotron-3B & -- & -- & -- & 0.7402 & 0.8462 & 0.7347 & 0.7350 & 0.8423 & 0.7303 & 0.7243 & 0.8327 & 0.7203 & 0.6475 & 0.7551 & 0.6459 \\
\bottomrule
\end{tabular}%
}
\end{table*}

\begin{table*}[htbp]
\centering
\caption{Comparison of ASR robustness on the FiQA dataset using \textbf{BM25} and \textbf{Qwen3-Embedding-8B} retrievers under different acoustic conditions. The metrics nDCG, MRR, and Recall denote nDCG@10, MRR@10, and Recall@10, respectively. Best results are highlighted in bold.}
\label{tab:fiqa_asr_merged}
\resizebox{\textwidth}{!}{%
\begin{tabular}{l ccc ccc ccc ccc}
\toprule
\multicolumn{13}{l}{\textbf{Dataset: FiQA}} \\
\midrule
\multirow{2}{*}{\textbf{ASR Model}} & \multicolumn{3}{c}{\textbf{Clean}} & \multicolumn{3}{c}{\textbf{Low (20dB)}} & \multicolumn{3}{c}{\textbf{Medium (10dB)}} & \multicolumn{3}{c}{\textbf{High (0dB)}} \\
\cmidrule(lr){2-4} \cmidrule(lr){5-7} \cmidrule(lr){8-10} \cmidrule(lr){11-13}
 & nDCG & MRR & Recall & nDCG & MRR & Recall & nDCG & MRR & Recall & nDCG & MRR & Recall \\
\midrule
\multicolumn{13}{c}{\textit{\textbf{Retrieval Model: BM25}}} \\
\midrule
Fun-ASR-Nano-2512 & 0.2180 & 0.2780 & 0.2656 & 0.2122 & 0.2712 & 0.2598 & 0.2129 & 0.2723 & 0.2604 & 0.2031 & 0.2584 & 0.2510 \\
GLM-ASR-Nano-2512 & 0.2141 & 0.2735 & 0.2657 & 0.2179 & 0.2771 & 0.2702 & 0.2171 & 0.2767 & 0.2693 & 0.2104 & 0.2655 & 0.2621 \\
Qwen3-ASR-1.7B & 0.2274 & 0.2870 & 0.2809 & 0.2204 & 0.2806 & 0.2705 & 0.2173 & 0.2757 & 0.2666 & 0.2150 & 0.2729 & 0.2653 \\
SenseVoice-Small & 0.2047 & 0.2616 & 0.2542 & 0.2011 & 0.2593 & 0.2486 & 0.1979 & 0.2556 & 0.2436 & 0.1891 & 0.2419 & 0.2366 \\
Whisper-Large-v3 & \textbf{0.2335} & \textbf{0.2946} & \textbf{0.2869} & \textbf{0.2329} & \textbf{0.2943} & \textbf{0.2887} & \textbf{0.2333} & \textbf{0.2934} & \textbf{0.2903} & \textbf{0.2214} & \textbf{0.2787} & \textbf{0.2736} \\
Whisper-Base & 0.2163 & 0.2755 & 0.2722 & 0.2179 & 0.2765 & 0.2745 & 0.2133 & 0.2703 & 0.2677 & 0.1902 & 0.2454 & 0.2373 \\
Whisper-Medium & 0.2278 & 0.2882 & 0.2819 & 0.2312 & 0.2918 & 0.2871 & 0.2309 & 0.2911 & 0.2866 & 0.2139 & 0.2723 & 0.2640 \\
Whisper-Small & 0.2268 & 0.2864 & 0.2818 & 0.2306 & 0.2913 & 0.2837 & 0.2277 & 0.2885 & 0.2827 & 0.2026 & 0.2555 & 0.2523 \\
Whisper-Tiny & 0.2129 & 0.2714 & 0.2642 & 0.2138 & 0.2711 & 0.2682 & 0.2031 & 0.2612 & 0.2499 & 0.1701 & 0.2192 & 0.2092 \\
\midrule
\multicolumn{13}{c}{\textit{\textbf{Retrieval Model: Qwen3-Embedding-8B}}} \\
\midrule
Fun-ASR-Nano-2512 & 0.5923 & 0.6574 & 0.6824 & 0.5873 & 0.6529 & 0.6741 & 0.5907 & 0.6562 & 0.6801 & 0.5689 & 0.6339 & 0.6536 \\
GLM-ASR-Nano-2512 & 0.5963 & 0.6689 & 0.6768 & 0.5930 & 0.6632 & 0.6762 & 0.5908 & 0.6617 & 0.6718 & 0.5700 & 0.6339 & 0.6548 \\
Qwen3-ASR-1.7B & \textbf{0.6030} & \textbf{0.6712} & \textbf{0.6899} & \textbf{0.5990} & \textbf{0.6689} & \textbf{0.6828} & 0.5959 & 0.6648 & \textbf{0.6808} & \textbf{0.5805} & \textbf{0.6426} & \textbf{0.6663} \\
SenseVoice-Small & 0.5732 & 0.6388 & 0.6592 & 0.5744 & 0.6390 & 0.6609 & 0.5716 & 0.6368 & 0.6548 & 0.5434 & 0.6038 & 0.6332 \\
Whisper-Large-v3 & 0.5968 & 0.6653 & 0.6793 & 0.5949 & 0.6647 & 0.6786 & \textbf{0.5974} & \textbf{0.6678} & 0.6788 & 0.5723 & 0.6357 & 0.6571 \\
Whisper-Base & 0.5727 & 0.6409 & 0.6560 & 0.5713 & 0.6369 & 0.6589 & 0.5686 & 0.6357 & 0.6538 & 0.4975 & 0.5628 & 0.5740 \\
Whisper-Medium & 0.5917 & 0.6617 & 0.6739 & 0.5920 & 0.6602 & 0.6778 & 0.5890 & 0.6551 & 0.6734 & 0.5589 & 0.6226 & 0.6423 \\
Whisper-Small & 0.5871 & 0.6581 & 0.6681 & 0.5854 & 0.6541 & 0.6669 & 0.5817 & 0.6479 & 0.6634 & 0.5343 & 0.5972 & 0.6124 \\
Whisper-Tiny & 0.5669 & 0.6360 & 0.6432 & 0.5670 & 0.6321 & 0.6503 & 0.5514 & 0.6191 & 0.6316 & 0.4636 & 0.5271 & 0.5264 \\
\bottomrule
\end{tabular}%
}
\end{table*}

\begin{table*}[htbp]
\centering
\caption{Comparison of ASR robustness on the HotpotQA dataset using \textbf{BM25} and \textbf{Qwen3-Embedding-8B} retrievers under different acoustic conditions. The metrics nDCG, MRR, and Recall denote nDCG@10, MRR@10, and Recall@10, respectively. Best results are highlighted in bold.}
\label{tab:hotpotqa_asr_merged}
\resizebox{\textwidth}{!}{%
\begin{tabular}{l ccc ccc ccc ccc}
\toprule
\multicolumn{13}{l}{\textbf{Dataset: HotpotQA}} \\
\midrule
\multirow{2}{*}{\textbf{ASR Model}} & \multicolumn{3}{c}{\textbf{Clean}} & \multicolumn{3}{c}{\textbf{Low (20dB)}} & \multicolumn{3}{c}{\textbf{Medium (10dB)}} & \multicolumn{3}{c}{\textbf{High (0dB)}} \\
\cmidrule(lr){2-4} \cmidrule(lr){5-7} \cmidrule(lr){8-10} \cmidrule(lr){11-13}
 & nDCG & MRR & Recall & nDCG & MRR & Recall & nDCG & MRR & Recall & nDCG & MRR & Recall \\
\midrule
\multicolumn{13}{c}{\textit{\textbf{Retrieval Model: BM25}}} \\
\midrule
Fun-ASR-Nano-2512 & 0.4846 & 0.6342 & 0.5118 & 0.4864 & 0.6362 & 0.5136 & 0.4840 & 0.6334 & 0.5113 & 0.4582 & 0.6036 & 0.4845 \\
GLM-ASR-Nano-2512 & 0.4821 & 0.6284 & 0.5123 & 0.4807 & 0.6269 & 0.5105 & 0.4814 & 0.6274 & 0.5117 & 0.4535 & 0.5930 & 0.4833 \\
Qwen3-ASR-1.7B & 0.5303 & 0.6904 & 0.5562 & 0.5187 & 0.6749 & 0.5457 & 0.5156 & 0.6706 & 0.5433 & 0.4958 & 0.6478 & 0.5228 \\
SenseVoice-Small & 0.4096 & 0.5412 & 0.4370 & 0.4065 & 0.5363 & 0.4350 & 0.4013 & 0.5292 & 0.4300 & 0.3768 & 0.4995 & 0.4030 \\
Whisper-Large-v3 & \textbf{0.5464} & \textbf{0.7098} & \textbf{0.5727} & \textbf{0.5426} & \textbf{0.7040} & \textbf{0.5697} & \textbf{0.5399} & \textbf{0.7008} & \textbf{0.5667} & \textbf{0.5128} & \textbf{0.6678} & \textbf{0.5394} \\
Whisper-Base & 0.4646 & 0.6147 & 0.4872 & 0.4628 & 0.6118 & 0.4866 & 0.4554 & 0.6019 & 0.4792 & 0.3978 & 0.5276 & 0.4207 \\
Whisper-Medium & 0.5237 & 0.6820 & 0.5500 & 0.5241 & 0.6833 & 0.5500 & 0.5194 & 0.6772 & 0.5459 & 0.4873 & 0.6380 & 0.5138 \\
Whisper-Small & 0.5013 & 0.6566 & 0.5255 & 0.5000 & 0.6553 & 0.5240 & 0.4961 & 0.6506 & 0.5203 & 0.4580 & 0.6040 & 0.4820 \\
Whisper-Tiny & 0.4314 & 0.5715 & 0.4554 & 0.4299 & 0.5699 & 0.4540 & 0.4168 & 0.5523 & 0.4415 & 0.3450 & 0.4617 & 0.3661 \\
\midrule
\multicolumn{13}{c}{\textit{\textbf{Retrieval Model: Qwen3-Embedding-8B}}} \\
\midrule
Fun-ASR-Nano-2512 & 0.6551 & 0.8147 & 0.6830 & 0.6520 & 0.8088 & 0.6818 & 0.6476 & 0.8040 & 0.6782 & 0.6113 & 0.7630 & 0.6426 \\
GLM-ASR-Nano-2512 & 0.6666 & 0.8223 & 0.6953 & 0.6657 & 0.8211 & 0.6945 & 0.6645 & 0.8196 & 0.6928 & 0.6253 & 0.7745 & 0.6552 \\
Qwen3-ASR-1.7B & 0.6802 & 0.8363 & 0.7082 & 0.6787 & 0.8364 & 0.7061 & 0.6771 & 0.8357 & 0.7034 & 0.6500 & 0.8046 & 0.6793 \\
SenseVoice-Small & 0.6115 & 0.7706 & 0.6421 & 0.6109 & 0.7700 & 0.6413 & 0.6049 & 0.7633 & 0.6356 & 0.5693 & 0.7210 & 0.6001 \\
Whisper-Large-v3 & \textbf{0.6968} & \textbf{0.8550} & \textbf{0.7235} & \textbf{0.6931} & \textbf{0.8514} & \textbf{0.7200} & \textbf{0.6891} & \textbf{0.8458} & \textbf{0.7167} & \textbf{0.6574} & \textbf{0.8120} & \textbf{0.6850} \\
Whisper-Base & 0.6292 & 0.7869 & 0.6579 & 0.6280 & 0.7862 & 0.6562 & 0.6154 & 0.7718 & 0.6433 & 0.5359 & 0.6762 & 0.5658 \\
Whisper-Medium & 0.6775 & 0.8361 & 0.7045 & 0.6770 & 0.8362 & 0.7038 & 0.6735 & 0.8321 & 0.7008 & 0.6308 & 0.7842 & 0.6578 \\
Whisper-Small & 0.6627 & 0.8220 & 0.6895 & 0.6593 & 0.8173 & 0.6869 & 0.6559 & 0.8144 & 0.6834 & 0.6037 & 0.7549 & 0.6324 \\
Whisper-Tiny & 0.5996 & 0.7573 & 0.6262 & 0.5947 & 0.7515 & 0.6219 & 0.5791 & 0.7340 & 0.6066 & 0.4738 & 0.6071 & 0.5024 \\
\bottomrule
\end{tabular}%
}
\end{table*}

\begin{table*}[htbp]
\centering
\caption{Comparison of ASR robustness on the NQ dataset using \textbf{BM25} and \textbf{Qwen3-Embedding-8B} retrievers under different acoustic conditions. The metrics nDCG, MRR, and Recall denote nDCG@10, MRR@10, and Recall@10, respectively. Best results are highlighted in bold.}
\label{tab:nq_asr_merged}
\resizebox{\textwidth}{!}{%
\begin{tabular}{l ccc ccc ccc ccc}
\toprule
\multicolumn{13}{l}{\textbf{Dataset: NQ}} \\
\midrule
\multirow{2}{*}{\textbf{ASR Model}} & \multicolumn{3}{c}{\textbf{Clean}} & \multicolumn{3}{c}{\textbf{Low (20dB)}} & \multicolumn{3}{c}{\textbf{Medium (10dB)}} & \multicolumn{3}{c}{\textbf{High (0dB)}} \\
\cmidrule(lr){2-4} \cmidrule(lr){5-7} \cmidrule(lr){8-10} \cmidrule(lr){11-13}
 & nDCG & MRR & Recall & nDCG & MRR & Recall & nDCG & MRR & Recall & nDCG & MRR & Recall \\
\midrule
\multicolumn{13}{c}{\textit{\textbf{Retrieval Model: BM25}}} \\
\midrule
Fun-ASR-Nano-2512 & 0.2772 & 0.2390 & 0.4315 & 0.2772 & 0.2391 & 0.4318 & 0.2746 & 0.2369 & 0.4274 & 0.2609 & 0.2259 & 0.4043 \\
GLM-ASR-Nano-2512 & 0.2805 & 0.2415 & 0.4368 & 0.2805 & 0.2414 & 0.4366 & 0.2795 & 0.2409 & 0.4342 & 0.2642 & 0.2274 & 0.4120 \\
Qwen3-ASR-1.7B & 0.2931 & 0.2521 & 0.4559 & 0.2891 & 0.2489 & 0.4492 & 0.2880 & 0.2478 & 0.4481 & 0.2763 & 0.2371 & \textbf{0.4324} \\
SenseVoice-Small & 0.2458 & 0.2098 & 0.3903 & 0.2493 & 0.2132 & 0.3937 & 0.2434 & 0.2078 & 0.3855 & 0.2259 & 0.1924 & 0.3605 \\
Whisper-Large-v3 & \textbf{0.2959} & \textbf{0.2547} & \textbf{0.4602} & \textbf{0.2954} & \textbf{0.2549} & \textbf{0.4579} & \textbf{0.2934} & \textbf{0.2529} & \textbf{0.4551} & \textbf{0.2780} & \textbf{0.2403} & 0.4291 \\
Whisper-Base & 0.2632 & 0.2256 & 0.4129 & 0.2612 & 0.2236 & 0.4107 & 0.2576 & 0.2211 & 0.4031 & 0.2192 & 0.1889 & 0.3424 \\
Whisper-Medium & 0.2880 & 0.2482 & 0.4475 & 0.2876 & 0.2483 & 0.4459 & 0.2851 & 0.2455 & 0.4434 & 0.2682 & 0.2317 & 0.4162 \\
Whisper-Small & 0.2793 & 0.2403 & 0.4342 & 0.2789 & 0.2399 & 0.4335 & 0.2780 & 0.2391 & 0.4324 & 0.2543 & 0.2201 & 0.3922 \\
Whisper-Tiny & 0.2456 & 0.2103 & 0.3857 & 0.2453 & 0.2107 & 0.3840 & 0.2358 & 0.2027 & 0.3702 & 0.1880 & 0.1616 & 0.2960 \\
\midrule
\multicolumn{13}{c}{\textit{\textbf{Retrieval Model: Qwen3-Embedding-8B}}} \\
\midrule
Fun-ASR-Nano-2512 & 0.6286 & 0.5791 & 0.8306 & 0.6252 & 0.5760 & 0.8261 & 0.6238 & 0.5749 & 0.8234 & 0.5890 & 0.5422 & 0.7806 \\
GLM-ASR-Nano-2512 & 0.6338 & 0.5850 & 0.8309 & 0.6332 & 0.5849 & 0.8299 & 0.6300 & 0.5820 & 0.8250 & 0.5956 & 0.5483 & 0.7871 \\
Qwen3-ASR-1.7B & 0.6388 & 0.5899 & 0.8385 & 0.6375 & 0.5877 & \textbf{0.8395} & 0.6342 & 0.5847 & \textbf{0.8352} & \textbf{0.6146} & \textbf{0.5668} & \textbf{0.8097} \\
SenseVoice-Small & 0.6126 & 0.5618 & 0.8162 & 0.6102 & 0.5593 & 0.8134 & 0.6058 & 0.5546 & 0.8093 & 0.5669 & 0.5174 & 0.7635 \\
Whisper-Large-v3 & \textbf{0.6415} & \textbf{0.5920} & \textbf{0.8430} & \textbf{0.6391} & \textbf{0.5901} & \textbf{0.8395} & \textbf{0.6350} & \textbf{0.5865} & 0.8342 & 0.6064 & 0.5592 & 0.8011 \\
Whisper-Base & 0.6046 & 0.5559 & 0.8023 & 0.5994 & 0.5506 & 0.7979 & 0.5905 & 0.5418 & 0.7893 & 0.5109 & 0.4677 & 0.6880 \\
Whisper-Medium & 0.6274 & 0.5783 & 0.8283 & 0.6297 & 0.5805 & 0.8297 & 0.6256 & 0.5761 & 0.8261 & 0.5897 & 0.5419 & 0.7837 \\
Whisper-Small & 0.6239 & 0.5742 & 0.8253 & 0.6240 & 0.5745 & 0.8248 & 0.6194 & 0.5702 & 0.8187 & 0.5662 & 0.5196 & 0.7554 \\
Whisper-Tiny & 0.5921 & 0.5431 & 0.7896 & 0.5868 & 0.5372 & 0.7857 & 0.5689 & 0.5204 & 0.7638 & 0.4623 & 0.4232 & 0.6239 \\
\bottomrule
\end{tabular}%
}
\end{table*}

\begin{table*}[htbp]
\centering
\caption{Comparison of ASR robustness on the DuRetrieval dataset using \textbf{BM25} and \textbf{Qwen3-Embedding-8B} retrievers under different acoustic conditions. The metrics nDCG, MRR, and Recall denote nDCG@10, MRR@10, and Recall@10, respectively. Best results are highlighted in bold.}
\label{tab:duretrieval_asr_merged}
\resizebox{\textwidth}{!}{%
\begin{tabular}{l ccc ccc ccc ccc}
\toprule
\multicolumn{13}{l}{\textbf{Dataset: DuRetrieval}} \\
\midrule
\multirow{2}{*}{\textbf{ASR Model}} & \multicolumn{3}{c}{\textbf{Clean}} & \multicolumn{3}{c}{\textbf{Low (20dB)}} & \multicolumn{3}{c}{\textbf{Medium (10dB)}} & \multicolumn{3}{c}{\textbf{High (0dB)}} \\
\cmidrule(lr){2-4} \cmidrule(lr){5-7} \cmidrule(lr){8-10} \cmidrule(lr){11-13}
 & nDCG & MRR & Recall & nDCG & MRR & Recall & nDCG & MRR & Recall & nDCG & MRR & Recall \\
\midrule
\multicolumn{13}{c}{\textit{\textbf{Retrieval Model: BM25}}} \\
\midrule
Fun-ASR-Nano-2512 & \textbf{0.5144} & \textbf{0.6453} & \textbf{0.5284} & \textbf{0.5115} & \textbf{0.6422} & 0.5255 & 0.5062 & \textbf{0.6374} & 0.5195 & 0.4842 & \textbf{0.6092} & 0.4992 \\
GLM-ASR-Nano-2512 & 0.5034 & 0.6313 & 0.5174 & 0.5029 & 0.6314 & 0.5158 & 0.4959 & 0.6233 & 0.5088 & 0.4534 & 0.5711 & 0.4695 \\
Qwen3-ASR-1.7B & 0.5108 & 0.6389 & 0.5255 & 0.5108 & 0.6377 & \textbf{0.5257} & \textbf{0.5077} & 0.6347 & \textbf{0.5222} & \textbf{0.4859} & 0.6086 & \textbf{0.5002} \\
SenseVoice-Small & 0.4724 & 0.5974 & 0.4868 & 0.4713 & 0.6000 & 0.4845 & 0.4649 & 0.5917 & 0.4788 & 0.4282 & 0.5457 & 0.4442 \\
Whisper-Large-v3 & 0.4401 & 0.5606 & 0.4533 & 0.4340 & 0.5542 & 0.4478 & 0.4245 & 0.5432 & 0.4373 & 0.3514 & 0.4521 & 0.3654 \\
\midrule
\multicolumn{13}{c}{\textit{\textbf{Retrieval Model: Qwen3-Embedding-8B}}} \\
\midrule
Fun-ASR-Nano-2512 & \textbf{0.8694} & \textbf{0.9217} & \textbf{0.8818} & \textbf{0.8684} & \textbf{0.9210} & \textbf{0.8805} & \textbf{0.8639} & 0.9164 & \textbf{0.8770} & 0.8297 & 0.8821 & 0.8438 \\
GLM-ASR-Nano-2512 & 0.8590 & 0.9140 & 0.8723 & 0.8568 & 0.9120 & 0.8704 & 0.8515 & 0.9065 & 0.8630 & 0.7912 & 0.8463 & 0.8060 \\
Qwen3-ASR-1.7B & 0.8645 & 0.9196 & 0.8763 & 0.8649 & 0.9203 & 0.8768 & 0.8615 & \textbf{0.9169} & 0.8727 & \textbf{0.8321} & \textbf{0.8877} & \textbf{0.8453} \\
SenseVoice-Small & 0.8565 & 0.9146 & 0.8692 & 0.8530 & 0.9110 & 0.8650 & 0.8462 & 0.9030 & 0.8575 & 0.7993 & 0.8550 & 0.8142 \\
Whisper-Large-v3 & 0.8036 & 0.8632 & 0.8165 & 0.7983 & 0.8597 & 0.8114 & 0.7880 & 0.8480 & 0.8002 & 0.6937 & 0.7535 & 0.7088 \\
\bottomrule
\end{tabular}%
}
\end{table*}

\begin{table*}[htbp]
\centering
\caption{Comparison of ASR robustness on the MedicalRetrieval dataset using \textbf{BM25} and \textbf{Qwen3-Embedding-8B} retrievers under different acoustic conditions. The metrics nDCG, MRR, and Recall denote nDCG@10, MRR@10, and Recall@10, respectively. Best results are highlighted in bold.}
\label{tab:medical_retrieval_asr_merged}
\resizebox{\textwidth}{!}{%
\begin{tabular}{l ccc ccc ccc ccc}
\toprule
\multicolumn{13}{l}{\textbf{Dataset: MedicalRetrieval}} \\
\midrule
\multirow{2}{*}{\textbf{ASR Model}} & \multicolumn{3}{c}{\textbf{Clean}} & \multicolumn{3}{c}{\textbf{Low (20dB)}} & \multicolumn{3}{c}{\textbf{Medium (10dB)}} & \multicolumn{3}{c}{\textbf{High (0dB)}} \\
\cmidrule(lr){2-4} \cmidrule(lr){5-7} \cmidrule(lr){8-10} \cmidrule(lr){11-13}
 & nDCG & MRR & Recall & nDCG & MRR & Recall & nDCG & MRR & Recall & nDCG & MRR & Recall \\
\midrule
\multicolumn{13}{c}{\textit{\textbf{Retrieval Model: BM25}}} \\
\midrule
Fun-ASR-Nano-2512 & 0.2848 & 0.2676 & 0.3400 & 0.2852 & 0.2679 & 0.3410 & \textbf{0.2835} & \textbf{0.2660} & \textbf{0.3400} & 0.2656 & 0.2488 & 0.3200 \\
GLM-ASR-Nano-2512 & 0.2679 & 0.2508 & 0.3230 & 0.2666 & 0.2492 & 0.3230 & 0.2650 & 0.2470 & 0.3230 & 0.2409 & 0.2235 & 0.2970 \\
Qwen3-ASR-1.7B & \textbf{0.2888} & \textbf{0.2713} & \textbf{0.3450} & \textbf{0.2866} & \textbf{0.2690} & \textbf{0.3430} & 0.2808 & 0.2633 & 0.3370 & \textbf{0.2697} & \textbf{0.2538} & \textbf{0.3210} \\
SenseVoice-Small & 0.2596 & 0.2443 & 0.3090 & 0.2557 & 0.2399 & 0.3070 & 0.2522 & 0.2368 & 0.3020 & 0.2351 & 0.2194 & 0.2860 \\
Whisper-Large-v3 & 0.2236 & 0.2080 & 0.2740 & 0.2147 & 0.2000 & 0.2620 & 0.2132 & 0.1975 & 0.2640 & 0.1807 & 0.1672 & 0.2240 \\
\midrule
\multicolumn{13}{c}{\textit{\textbf{Retrieval Model: Qwen3-Embedding-8B}}} \\
\midrule
Fun-ASR-Nano-2512 & 0.5910 & 0.5598 & 0.6910 & \textbf{0.5945} & 0.5624 & \textbf{0.6970} & \textbf{0.5962} & \textbf{0.5642} & \textbf{0.6980} & \textbf{0.5724} & \textbf{0.5400} & \textbf{0.6760} \\
GLM-ASR-Nano-2512 & 0.5741 & 0.5428 & 0.6740 & 0.5788 & 0.5476 & 0.6790 & 0.5749 & 0.5430 & 0.6770 & 0.5262 & 0.4944 & 0.6290 \\
Qwen3-ASR-1.7B & \textbf{0.5949} & \textbf{0.5634} & 0.6960 & 0.5935 & \textbf{0.5632} & 0.6910 & 0.5882 & 0.5577 & 0.6860 & 0.5666 & 0.5348 & 0.6690 \\
SenseVoice-Small & 0.5898 & 0.5562 & \textbf{0.7010} & 0.5854 & 0.5520 & 0.6960 & 0.5817 & 0.5495 & 0.6880 & 0.5583 & 0.5266 & 0.6630 \\
Whisper-Large-v3 & 0.5301 & 0.4985 & 0.6300 & 0.5285 & 0.4968 & 0.6290 & 0.5299 & 0.4974 & 0.6330 & 0.4673 & 0.4336 & 0.5750 \\
\bottomrule
\end{tabular}%
}
\end{table*}

\begin{table*}[htbp]
\centering
\caption{Comparison of ASR robustness on the T2Retrieval dataset using \textbf{BM25} and \textbf{Qwen3-Embedding-8B} retrievers under different acoustic conditions. The metrics nDCG, MRR, and Recall denote nDCG@10, MRR@10, and Recall@10, respectively. Best results are highlighted in bold.}
\label{tab:t2retrieval_asr_merged}
\resizebox{\textwidth}{!}{%
\begin{tabular}{l ccc ccc ccc ccc}
\toprule
\multicolumn{13}{l}{\textbf{Dataset: T2Retrieval}} \\
\midrule
\multirow{2}{*}{\textbf{ASR Model}} & \multicolumn{3}{c}{\textbf{Clean}} & \multicolumn{3}{c}{\textbf{Low (20dB)}} & \multicolumn{3}{c}{\textbf{Medium (10dB)}} & \multicolumn{3}{c}{\textbf{High (0dB)}} \\
\cmidrule(lr){2-4} \cmidrule(lr){5-7} \cmidrule(lr){8-10} \cmidrule(lr){11-13}
 & nDCG & MRR & Recall & nDCG & MRR & Recall & nDCG & MRR & Recall & nDCG & MRR & Recall \\
\midrule
\multicolumn{13}{c}{\textit{\textbf{Retrieval Model: BM25}}} \\
\midrule
Fun-ASR-Nano-2512 & \textbf{0.5297} & \textbf{0.6765} & \textbf{0.5212} & \textbf{0.5289} & \textbf{0.6758} & \textbf{0.5203} & \textbf{0.5271} & \textbf{0.6743} & \textbf{0.5185} & \textbf{0.5040} & \textbf{0.6470} & 0.4963 \\
GLM-ASR-Nano-2512 & 0.5201 & 0.6654 & 0.5123 & 0.5196 & 0.6651 & 0.5116 & 0.5159 & 0.6611 & 0.5084 & 0.4772 & 0.6143 & 0.4711 \\
Qwen3-ASR-1.7B & 0.5253 & 0.6710 & 0.5171 & 0.5242 & 0.6699 & 0.5160 & 0.5220 & 0.6677 & 0.5140 & 0.5037 & 0.6457 & \textbf{0.4966} \\
SenseVoice-Small & 0.4950 & 0.6395 & 0.4884 & 0.4912 & 0.6346 & 0.4846 & 0.4869 & 0.6298 & 0.4802 & 0.4568 & 0.5933 & 0.4518 \\
Whisper-Large-v3 & 0.4517 & 0.5888 & 0.4480 & 0.4497 & 0.5870 & 0.4456 & 0.4391 & 0.5754 & 0.4348 & 0.3739 & 0.4922 & 0.3722 \\
\midrule
\multicolumn{13}{c}{\textit{\textbf{Retrieval Model: Qwen3-Embedding-8B}}} \\
\midrule
Fun-ASR-Nano-2512 & \textbf{0.8405} & \textbf{0.9074} & \textbf{0.8324} & \textbf{0.8395} & \textbf{0.9061} & \textbf{0.8317} & \textbf{0.8365} & \textbf{0.9039} & \textbf{0.8283} & 0.8032 & 0.8710 & 0.7961 \\
GLM-ASR-Nano-2512 & 0.8246 & 0.8924 & 0.8185 & 0.8240 & 0.8920 & 0.8177 & 0.8185 & 0.8867 & 0.8123 & 0.7648 & 0.8328 & 0.7607 \\
Qwen3-ASR-1.7B & 0.8370 & 0.9040 & 0.8287 & 0.8360 & 0.9031 & 0.8281 & 0.8336 & 0.9006 & 0.8258 & \textbf{0.8067} & \textbf{0.8738} & \textbf{0.8003} \\
SenseVoice-Small & 0.8222 & 0.8918 & 0.8155 & 0.8189 & 0.8886 & 0.8125 & 0.8149 & 0.8852 & 0.8088 & 0.7752 & 0.8453 & 0.7711 \\
Whisper-Large-v3 & 0.7719 & 0.8460 & 0.7685 & 0.7694 & 0.8437 & 0.7666 & 0.7607 & 0.8350 & 0.7587 & 0.6775 & 0.7509 & 0.6780 \\
\bottomrule
\end{tabular}%
}
\end{table*}

\end{document}